\documentclass[12pt]{article}

 \usepackage{setspace}% \doublespacing

\usepackage{natbib} % biblio
\usepackage{amsmath}
\usepackage{amsfonts}
\usepackage{amssymb,tabularx}
\usepackage{enumerate,color}

\usepackage{graphicx,subfigure,placeins}

\renewcommand{\thefigure}{\arabic{figure}}
\renewcommand{\thetable}{\arabic{table}}

\def\r{\mathbf r}
\def\U{\mathbf U}
\def\C{\mathbf C}
\def\W{\mathbf W}
\def\X{\mathbf X}
\def\E{\mathbf E}
\def\e{\mathbf e}
\def\V{\mathbf V}
\def\R{\mathbb R}

\def\Z{\mathbf Z}
\def\P{\mathbf P}
\def\D{\mathbf D}
\def\b{\boldsymbol \beta}
\def\a{\boldsymbol \alpha}\def\g{\boldsymbol \gamma}
\def\S{\boldsymbol \Sigma}
\def\bxi{\boldsymbol \xi}

\newtheorem{definition}{{\bf Definition}}
\newtheorem{theorem}{{\bf Theorem}}
\newtheorem{corollary}{{\bf Corollary}}
\newtheorem{proposition}{{\bf Proposition}}

\begin{document}

\vskip 0.22in \centerline{\large\bf  Survival Model Construction}
  \centerline{\large\bf Guided by Fit and Predictive Strength}

\vskip 0.252in \vskip 0.152in \centerline{\bf C\'ecile Chauvel and John O'Quigley}

\vskip 0.252in 

\centerline{\rm Laboratoire de Statistique Th\'eorique et Appliqu\'ee,}
\centerline{\rm Universit\'e Pierre et Marie Curie - Paris VI, 75005 Paris, France}

\vskip 0.252in \vskip 0.152in 

\noindent{\bf ABSTRACT}: 
We describe a unified framework within which we can build survival models. The motivation for this work comes from a study on the prediction of relapse among breast cancer patients treated at the Curie Institute in Paris, France. 
Our focus is on how to best code, or characterize, the effects of the variables, either alone or in combination with others. 
We consider simple graphical techniques that not only provide an immediate indication as to the goodness of fit but, in cases of departure from model assumptions, point in the direction of a more involved alternative model. 
These techniques help support our intuition. This intuition is backed up by formal theorems that underlie the process of building richer models from simpler ones. Goodness--of--fit techniques are used alongside measures of predictive strength and, again, formal theorems show that these measures can be used to help identify models closest to the unknown non--proportional hazards mechanism that we can suppose generates the observations.
We consider many examples and show how these tools can be of help in guiding the practical problem of efficient model construction for survival data. 
 \par

\vskip 0.19in {\noindent \it Key words}: \ Proportional Hazards; Goodness of fit; Predictive measures; $R^2$ measures; Time--varying coefficient.

\section{INTRODUCTION}
\subsection{Motivation}
The advent of personalized medicine together with rapid progress in techniques of genetics, next generation sequencing for example, the use of biomarkers, together with analytic techniques in bioinformatics have brought a renewed focus on the problems of model--based prediction. 
The related but different question concerning goodness of fit for any model has been given rather greater attention, at least in the survival literature. 
$R^2$ measures quantify the predictive capacity of a model, and this may be high even when the model assumptions are seriously violated, whereas goodness--of--fit measures focus on the model assumptions and aim to examine how well these are supported by the data themselves. Although a number of authors have carefully outlined that distinction it is true that some confusion still remains. \newline

Our motivation stems from a study of $1504$ breast cancer patients treated at the Institut Curie in Paris, France. Subsequent to initial treatment, patients were followed for a period of fifteen years. Among several study objective relating to this cohort was the aim to construct descriptive survival models that could provide a deeper understanding to prognosis after initial treatment. The problem is inherently a multi--factorial one. Combined effects of prognostic factors as well as conditional effets are a central concern. By conditional, we mean the impact of various risk factors on survival after having taken account of the impact of known or suspected risk factors. For instance, it can be of interest to try to quantify the added prognostic information of a more or less complex construction of biomarkers after having already accounted for known clinical risk factors. Finally, the effect of several of these risk factors can change with time and useful prognostic indices should reflect such time dependencies.

\subsection{Background}

Goodness--of--fit procedures can be directed at more than one aspect of any model. We may wish to consider overall fit of the model, i.e., how well the model when taken as a whole is supported by the observations or we may wish to focus on some particular feature of the model and how well it holds up in practice. For example, we may be interested in checking the working assumptions regarding treatment differences in presence of other covariates when the model fit of these covariates is of only indirect concern. 
The goodness of fit of a model can be evaluated by using tests or graphical methods. In this paper, we focus on graphical methods that can not only indicate departures from working assumptions but can also, of themselves, suggest remedies. 
The first graphical method for checking the proportional hazards assumption was proposed by \citet{Kay1977} who suggested to plot an estimate of the conditional cumulative hazard $\Lambda(t \vert Z)$ over time. When $Z$ is a categorical covariate, typically representing treatment groups, the plot should result in parallel curves under proportional hazards. \citet{Andersen1982b} extended this approach to continuous covariates by discretizing them. 
Other graphical methods based on residuals can be sorted in two categories, depending on whether the residuals are cumulated or not. Amongst non-cumulative methods, a large class of martingale residuals described by \citet{Barlow1988} can be used by plotting their members over time. The \citet{Schoenfeld1982} residuals, weighted Schoenfeld residuals introduced by \citet{Lin1993} and the residuals of \citet{Kay1977} arise as special members of this class. \citet{Grambsch1994} suggested plotting standardized residuals over time to detect the validity of proportional hazards assumption and, in case of rejection, have an indication on the shape of the time-varying effect. This is the most commonly used approach and is implemented in the programming languages R and Splus.
More recently, \citet{Sasieni2003} proposed the use of martingale difference residuals. The latter method requires care in interpretation since several plots corresponding to several time points have to be considered. All of these non--cumulative residual methods presented so far make use of a smoothing function to average the residual points. As pointed out by \citet{Lin1993}, the result can be sensitive to the choice of the smoothing techniques. 
To overcome this problem, several authors proposed the use of cumulative martingale residuals, such as \citet{Arjas1988}, 
\citet{Therneau1990} and \citet{Lin1993}. The method of \citet{Therneau1990} is based on the score process of \citet{Wei1984}. 
Under the proportional hazards assumption, this process converges weakly to a Brownian bridge and a test of the supremum of a Brownian bridge can be performed. \citet{Lin1993} showed that Wei's score process can be asymptotically approximated by a gaussian process with a data--based variance--covariance matrix. Therefore, the comparison between the observed score process and a large numbers of simulated outcomes of the limiting gaussian process can give an indication of the validity of the proportional hazards assumption. In practice, the interpretation of such a plot is not always clear. 
More details about goodness--of--fit methods can be found in \citet{Klein2003}, \citet{Therneau2000} and more recently \citet{Martinussen2005}.\\

Unlike the case of linear regression, if the multivariate proportional hazards model holds, the sub--models will no longer be simultaneously valid. Therefore, the evaluation of the goodness of fit of the multivariate model by evaluating the fit of the univariate sub-models will not suffice. However, in absence of tools for checking the overall validity of the model, most of the existing methods for checking the fit of one covariate assume proportional hazards for the other covariates, which is an erroneous assumption (\citealt{Scheike2004}). Besides, the validity of the results of such methods depends on the covariance between covariates. To adress this issue, \citet{Scheike2004} considered a non--proportional hazards model and developed estimation procedures and tests of the goodness of fit for one covariate with the possibility for the others not to have a constant regression effect. Their simulation work indicates the good performance of their method when compared to several existing and commonly used methods 
when the proportional hazards assumption is not met and/or in the presence of correlated covariates. Their test statistic depends on the estimation of the regression parameter requiring an involved algorithm relying on kernel estimation. The shape of the resulting estimator of the regression parameter is not an explicit and smooth function of time. The expression of the asymptotic distribution is unavailable for their statistic.   
The goodness--of--fit evaluation procedure presented in this article is a graphical method which does not require any estimation and is simple to understand. Our method is also based on the general framework of a non--proportional hazards model and is adapted to multivariate settings with correlated covariates. 
\\

Measures of predictive ability, on the other hand, - we will focus specifically on $R^2$ type measures - are used to examine several different questions. Typical questions may be, how well does some set of biological markers perform, in a predictive sense, when compared to some other set. How much added predictive information is contained in a biomarker when added to already known clinical prognostic factors such as stage and grade. When all known factors are included in a model, how much of the variability is accounted for so that, in consequence, how much variability remains to be explained, either by physical or possibly genetic attributes. Finally, how does the relaxing of certain model assumptions - one example would be stratification rather than inclusion in the linear component of a proportional hazards model - impact prediction. This last observation draws attention to the fact that, although different techniques with a different purpose, the aims of goodness--of--fit procedures and predictive measures can to some degree overlap.
In the context of survival analysis, in particular when using the Cox proportional hazards model, several authors have proposed different measures of predictive ability.
 A recent and exhaustive literature review on the predictive accuracy measures can be found in \citet{Choodari2012}.
 No consensus has yet been established regarding the most suitable measure to use in practice (\citealt{Muller2008}, \citealt{Hielscher2010}, \citealt{Choodari2012}). \newline

It is not clear in what way, or in what sense, an improvement in predictability implies an improvement in goodness of fit. In fact it is not difficult to come up with counter examples and the notion itself is not very precise. The converse is however correct, and, in this work, we prove in a theorem that an improvement in fit of a proportional hazards model results in an improvement in predictability. This theorem underlies the purpose of this article which is 
 to investigate ways to improve goodness--of--fit for proportional hazards type models and to see how this impacts the resulting predictive power of the model. We work with goodness--of--fit procedures and measures of predictive ability that are closely related, having as their basis the residuals from the non--proportional hazards model. The goodness of fit is evaluated with a version of the score process introduced by O'Quigley  (\citeyear{OQuigley2003}, \citeyear{OQuigley2008} chap. 8) which is extended here to the multivariate setting. We obtain the exact expression of the limiting distribution of the process. The predictive accuracy measure is the $R^2$ coefficient described by \citet{OQuigley1994} but is also extended to the multivariate non--proportional hazards situation. 
  This leads to easily assessed visual techniques and provides a complete and unified approach to the testing, fit and quantification of predictive effects. Several examples illustrate the ideas.

In the next section we describe the non--proportional hazards model and use it to derive stochastic processes of particular relevance to the problem we are studying. In Section 3, we present the main result that indicates why improvements in fit will result in improvements in predictive capability and how to proceed in practice. Section 4 summarizes simulations that provide additional support to our intuition and an application to a real dataset is provided. Before that, we recall the main notation.

\subsection{Notation} 
The random variables of interest are the failure times $T_i$, the
censoring times $C_i$ and the vector of dimension $p$ of possibly time-dependent covariates
$\Z_i=(Z_i^1,\dots,Z_i^p)$, $i=1,...,n.$ We view these as a random sample from
the distribution of $T$, $C$ and $\Z=(Z^1,\dots,Z^p)$ which
have support on some finite interval. To emphasis the time-dependence, with a slight abuse of notation, we refer to any time-dependent quantity $A$ as $A(t)$, $A$ being either random or deterministic.  
 The time-dependent covariate $\Z(t) $ is assumed to be a predictable stochastic process which admits a moment of order 4. 
For each subject $i$, the observed time is  $X_i=\min(T_i, C_i)$, and the observed indicator of failure is
$\delta_i= I(T_i\leq C_i)$, where $I$ is the indicator function. The at--risk indicator $Y_i(t)$ is
defined as $Y_i(t)= I(X_i\geq t).$ The counting process $N_i(t)$
is defined as $N_i(t)=I(T_i\leq t, T_i\leq C_i)$ and we also
define $\bar{N}(t)=\sum_{i=1}^n N_i(t)$. 
It is of
notational convenience to define ${\cal Z }(t)= \sum_{i=1}^n
\Z_i(t)I(X_i=t,\delta_i=1)$, in words a $\R^p$-valued function equal
to zero except at the observed failures where it assumes
the covariate value of the subject that fails.
In addition, $\|\mathbf{a}\|=\underset{i=1,\dots,p}{\max}|a_i| $ denotes the maximum  norm of the vector $\mathbf{a}=(a_1,\dots,a_p) \in \R^p$. For a $p\times p$ matrix $\mathbf{A}$ with element $(i,j)$ denoted $A_{i,j}$, $i,j=1,\dots,p$, $\|\mathbf{A}\|=\underset{i,j=1,\dots,p}{\max}|A_{i,j}| $ denotes the maximum 
 norm of $\bf A$. Let ${\bf A}^T$ (respectively ${\bf a}^T$) denote the transpose of the matrix $\mathbf{A}$ (resp. vector $a$). The product $\mathbf{a}^{\otimes 2}=\mathbf{aa}^T$ is the matrix with element $[\mathbf{a}^{\otimes 2}]_{i,j}=a_ia_j$. Denote $\det(\mathbf{A})$ the determinant of the matrix $\bf A$. The space $D[0,1]^p=D[0,1]\times \dots \times D[0,1]$ is equipped with the Skorokhod product topology.

\section{MODEL-BASED EMPIRICAL PROCESSES}

Consider the non--proportional hazards model defined by
\begin{equation}
\label{nonph}
\lambda\left\{t\mid\Z(t)\right\}=\lambda_0(t)\exp\left\{\b(t)^T \Z(t)\right\}, 
\end{equation}
 where
$\lambda(t|\cdot)$ is the conditional hazard function,
$\lambda_0(t)$ is a baseline hazard, $\b(t)$ is the time--dependent
regression effect and has dimension $p$ and  $\b(t)^T \Z(t)$ is the usual inner product between $\b(t)$ and $\Z(t)$. This model has been considered previously by several authors \citep{Murphy1991,Hastie1990,Zucker1990,Cai2003,Winnett2003,Scheike2004}.
With covariates constant over time,
the above model becomes  the proportional hazards model \citep{Cox1972} under the restriction that $\b(t)=\b$.
When we take the risk sets to be fixed and known and conditional on a failure at time $t$, the probability that the failure concerns individual $i$ is 
\begin{equation}
\label{2.1} \pi_i(\b(t),t) =  { Y_i(t) \exp\{
\b(t)^T \Z_i(t) \} }/\sum_{j=1}^n Y_j(t)\exp\{ \b(t)^T \Z_j(t) \}
,\quad i=1,\dots,n.
\end{equation} 
The expectation and variance with respect to the probabilities $\{\pi_i(\beta(t),t)\}_{i=1,\dots,n}$ are respectively a vector $\E_{\b(t)}\left(Z\vert t\right)$ of dimension $p$ and a $p\times p$ matrix $\V_{\b(t)}\left(Z\vert t\right)$ such that, 
\begin{align*}
\E_{\b(t)}\left(Z\vert t\right) &= \sum_{i=1}^n  Z_i(t) \pi_i( \b(t),t), \\
\V_{\b(t)}\left(Z\vert t\right) &= \sum_{i=1}^n  Z_i(t)^{\otimes 2} \pi_i( \b(t),t) -\E_{\b(t)}\left(Z\vert t\right)^{\otimes 2}.
\end{align*}
These quantities correspond to the conditional moments of the process $\mathcal{Z}(t)$ for a fixed $t$, given the risk sets.
The conditional variance-covariance matrix $\V_{\b(t)}\left(Z\vert t\right)$ is symmetric and positive definite. Thus, there exists an orthogonal matrix $\P_{\b(t)}(t)$ and a diagonal matrix $\D_{\b(t)}(t)$ such that 
 \begin{equation*}
\V_{\b(t)}\left(Z\vert t\right)=\P_{\b(t)}(t)\D_{\b(t)}(t)\P_{\b(t)}(t)^T.
 \end{equation*}
This leads us to define the symmetric matrix $\V_{\b(t)}\left(Z\vert t\right)^{x}$ by 
 \begin{equation*}
\V_{\b(t)}\left(Z\vert t\right)^{x}=\P_{\b(t)}(t)\left(\D_{\b(t)}(t)\right)^{x}\,\P_{\b(t)}(t)^T,\quad x\in\{-1/2,1/2\}.
 \end{equation*}

Denote \begin{equation}
\mathbf{r}_{\b(t)}(t)={\cal Z}(t)-\E_{\b(t)}(Z|t), \label{schoenfeldres}
\end{equation}
the residuals of the non--proportional hazards model (\ref{nonph}) with parameter $\b(t)$ evaluated at time $t$. In the case of the proportional hazards model, these residuals reduce to the well-known Schoenfeld residuals (\citealt{Schoenfeld1982}). Assume the case of a unique covariate ($p=1$) resulting in a univariate regression coefficient $\beta(t)$, a univariate conditional expectation $E_{\beta(t)}(Z\vert t)$ and a univariate residual $r_{\beta(t)}(t)$. Consider the partial scores 
\begin{equation}
\label{gof1} U(\beta(t),t) = \int_0^t r_{\beta(s)}(s)% w(s) 
d\bar N(s).
\end{equation}  
With a constant regression effect $\beta$, these correspond to the partial scores of \citet{Wei1984}. 
Wei was interested in goodness of fit for the two group problem and based a test on $\sup_t |
U(\hat\beta,t)|$, large values indicating departures away from
proportional hazards in the direction of non--proportional
hazards. Considerable exploration of this idea, and substantial
generalization via the use of martingale--based residuals, has
been carried out by
\citet{Lin1993,Lin1996} who showed that a wide choice of functions, potentially describing different
kinds of departures from the model could be used. Apart from the
two--group case, limiting distributions are complicated and
usually approximated via simulation. Furthermore, \citet{Lin1993} pointed out that extensions of their methodology to the multivariate case or to the integration of time--dependent covariates are not straightforward.
In order to overcome these difficulties, we follow the construction developed by  \citet{Khmaladze1981}, working directly with the increments of the process
rather than the process itself. We are then able to derive related processes
for which the limiting distributions are available analytically. To be more specific, when working with the ranks of the failure times and standardizing each increment of the process with a particular value rather than applying the same standardization for the whole score process, the limiting distribution of the multivariate process can be anticipated analytically and time-dependent covariates can be directly taken into account.

\subsection{Time Scale}
Let $k_n=\#\{i : i=1,\dots,n,\ \delta_i=1, \, \det\left (\V_{\b(X_i)}(Z|X_i)\right)>0\}$, where $\#A$ denotes the cardinality of the set $A$, denote the number of observed failures such that the conditional variances assessed at the event-times are positive--definite matrices. 
 In our setting, a null conditional variance at any time implies null conditional variances at later times. We assume that the number of failures $k_n$ increases without bound as $n$ increases without bound. By virtue of the fact that in Equation (\ref{nonph}), $\lambda_0(t)$ is unspecified, a monotonically increasing transformation of the times leaves inference for the regression parameter of the proportional hazards model unchanged. Therefore, \citet{Chauvel2014} considered the transformed times $\phi_n(X_i)$ such that
\begin{equation}\label{transfo_phi}
\phi_n(X_i)= \dfrac{ \bar N(X_i) }{k_n}\left(1+ (1-\delta_i) \dfrac{\# \left\lbrace j:j=1,\dots,n, X_j<X_i,\ \bar N(X_j)=\bar N(X_i) \right\rbrace}{ \#\left\lbrace j:j=1,\dots,n, \bar N(X_j)=\bar N(X_i)\right\rbrace}\right).
\end{equation}
Recall that the counting process $\{\bar N(t)\}_{t \in \mathcal{T} }$ presents a unit jump at each observed failure time. On the new scale, the times in the set $\{0,1/k_n,2/k_n,\dots,1\}$ correspond to failure times, the $i$th ordered failure time, denoted $t_i$, is such that $t_i=i/k_n$. The set $\{0,1/k_n,2/k_n,\dots,1\}$ is included in but not equal to the set of images of all failure times. 
Censoring times can assume any value as long as they keep their original locations between adjacent failure times. For simplicity in Formula (\ref{transfo_phi}), we take these times to be spread uniformly between adjacent failure times, maintaining the original ranking. The time $t_0$ on this scale corresponds to the $100\times t_0$th percentile of failure in the sample. For instance, at time $t_0=0.5$, half of the failures are observed. The inverse transformation of $\phi_n$ can be easily obtained and would enable us to interpret the results on the original time scale. 
On this transformed time scale, we can define the at-risk indicator $ Y_i^*(t) $ by $ Y_i^*(t)=I(\displaystyle \phi_n(X_i)\geq t)$ and the individual counting process $ N_i^*(t) =I(\displaystyle \phi_n(X_i) \leq t, \delta_i=1)$, for individual $i=1,\dots,n$. In what follows, we only work with the standardized time scale, so that the process $\mathcal{Z}$, the expectation $\E_{\b(t)}(Z|t)$ and the variance $\V_{\b(t)}(Z|t)$, of which extensions are straightforward, are defined for $0\leq t\leq 1$. 
 Define the counting process associated with the transformed times which has unit jumps at failure--times on the new scale by  $$ \bar N^* (t)=\sum_{i=1}^n I( \phi_n(X_i) \leq t, \delta_i=1),\quad 0\leq t\leq 1.$$ 
On the new time scale, the 
partial scores (\ref{gof1}) can be re-expressed as
 \begin{equation*}
  \U(\beta(t),t) =\int_0^{t} \r_{\b(s)}(s)
  d\bar N^*(s)=\sum_{i=1}^{\lfloor k_n t \rfloor} 
  \r_{\b(t_i)}(t_i)
 , \quad \quad 0\leq t \leq 1,
 \end{equation*}
where the $i$th element of the vector $\int_0^t\mathbf{a}(s)d\bar N^*(s)$ is $\int_0^t\mathbf{a}_i(s)d\bar N^*(s)$ for any $\R^p$-valued $\mathbf{a}(t)=(a_1(t),\dots,a_p(t))$, $i=1,\dots,p$ and 
$\lfloor x \rfloor$ gives the largest integer less than or equal to $x$.

\subsection{Multivariate Standardized Score Process}
\label{sec:process}
Before defining the standardized score process, let us give the assumptions needed in the sequel.
Let  $t \in [0,1]$, $\g(t)$ be a regression function, not necessarily equals to $\b(t)$ and
\begin{align*}
S^{(0)}(\g(t),t) = n^{-1} \sum^n_{i=1}
Y_i(t)&e^{\g(t) Z_i(t)}, \quad \quad 
\mathbf{S}^{(1)}(\g(t),t) = n^{-1} \sum^n_{i=1}
Y_i(t)Z_i(t)e^{\g(t) Z_i(t)}, \\
\mathbf{S}^{(2)}(\g(t),t) &= n^{-1} \sum^n_{i=1}
Y_i(t)Z_i(t)^{\otimes 2}e^{\g(t) Z_i(t)}.
\end{align*} 
Using these notations, we have the equalities 
$\E_{\g(t)}(Z\vert t)=\mathbf{S}^{(1)}(\g(t),t)/S^{(0)}(\g(t),t)$ 
 and 
$\V_{\g(t)}(Z \vert t)=\mathbf{S}^{(2)}(\g(t),t)/S^{(0)}(\g(t),t) -\E_{\g(t)}(Z\vert t)^{\otimes 2}$. Notice that the Jacobian matrix of $\E_{\g(t)}(Z\vert t)$ is the variance--covariance matrix $\V_{\g(t)}(Z \vert t)$. Consider that the following assumptions, similar to those of \citet{Andersen1982} hold:
\begin{enumerate}[A.]
\item \label{point1_multiv} (Asymptotic stability). There exists a neighbourhood $\mathcal{B}$ of $\b(t)$ and vector and matrix functions $\mathbf{s}^{(r)}(\g(t),t)$, $r=0,1,2$, defined for $t\in[0,1]$ and $\g(t)\in \mathcal{B} $ such that $\bf 0$ and $\b(t)$ are in the interior of $\mathcal{B}$, for all $t \in [0,1]$ and
\begin{eqnarray*}
\sqrt n \sup_{t\in[0,1],\g(t)\in\mathcal{B}} \left\| \mathbf{S}^{(r)}(\g(t),t)-\mathbf{s}^{(r)}(\g(t),t)\right\| \underset{n \rightarrow \infty}{\overset{ \mathbb{P}}{\longrightarrow}} 0 .
\end{eqnarray*}
\item (Asymptotic regularity). \label{point2_multiv} All functions defined in assumption \ref{point1_multiv}. are uniformly continuous in $t\in [0,1]$. In addition, for $r=0,1,2$, $s^{(r)}(\g(t),t)$ are continuous functions of $\g(t) \in \mathcal{B}$, bounded on $\mathcal{B}\times [0,1]$ and $s^{(0)}(\g(t),t)$ is bounded away from $0$.
\item (Homoscedasticity). \label{homosced_multiv}
There exists a symmetric and positive definite matrix $\S$ and a series of positive constants $(M_n)_n$ converging to $0$ as $n$ goes to infinity such that 
\begin{eqnarray*}
 \sup_{t\in[0,1],\g(t)\in\mathcal{B}} \left\| \left.\dfrac{\partial}{\partial \b } \V_{\b}(Z \vert t)\right\vert_{\b=\g(t)}  \right\| &\leq& M_n \quad a.s.,
\\
\sup_{t\in[0,1],\g(t)\in\mathcal{B}}   \left\|  \V_{\g(t)}(Z \vert t) - \S  \right\| &\underset{n \rightarrow \infty}{\overset{\bf L^1}{\longrightarrow}}& 0. 
\end{eqnarray*}
\end{enumerate} 
By analogy with the empirical quantities, we denote $e(\g(t),t)=s^{(1)}(\g(t),t)/s^{(0)}({\bf 0},t)$ and 
$v(\g(t),t)=s^{(2)}(\g(t),t)/s^{(0)}({\bf 0},t)-e(\g(t),t)^{\otimes 2}$.

The two first conditions are classical and introduced by \citet{Andersen1982} for using counting process and martingales theory, such as Lenglart's inequality or Rebolledo's theorem. Although we use a different approach here, that we believe is simpler to comprehend, the same assumptions are made.
 Notice that $\V_{\g(t)}(Z \vert t)$ is, by definition, the sample-based variance of $Z$ given $T=t$ under the model with parameter $\g(t)$. Thus, condition \ref{homosced_multiv}. of homoscedasticity means that the asymptotic variance does not depend on time. This condition is implicitly used in the context of the proportional hazards regression, for instance when estimating the variance of the parameter $\b$ or when applying the log-rank test. The contribution to the global variance is the same at each failure time, by the use of an unweighted sum of each term. 
 This stability of variance has also been noticed by several authors, for example \citet{Grambsch1994}. 
From the previous section, under the non--proportional hazards model (\ref{nonph}), the increments of the process $ \sum_{j=1}^{\lfloor k_n t \rfloor} \mathcal{Z}(t_j) $ at $t=t_i$ have mean ${\E}_{\b(t_i)}(Z|t_i)$ and
variance-covariance matrix ${\V}_{\b(t_i)} \left(Z\vert t_i\right).$  The increments of the process are
independent, either by design in view of the conditional model, or by the arguments of  \citet{Cox1975}. Thus only the
existence of the variance is necessary to carry out
appropriate standardization and to be able to appeal to the
functional central limit theorem. This leads us to define a standardized version of the multivariate score process:
%%%%%%%%%%%%%%%%%%%%%%%%%%%%%%%%%%%%%%%%%%%%%%%%%%%%%%%%%%%%%%%%%%%%%%%%%%%%
\begin{definition}
The multivariate standardized score process evaluated at parameter $\b_0$ and at failure time $t\in\{ 0,1/k_,2/k_n\dots,1\}$ is 
\begin{equation*}
\U^*(\b_0,t)=\dfrac{1}{\sqrt{k_n}} \int_0^{t} \V_{\b_0} \left(Z\vert s\right)^{-1/2}\r_{\b_0}(s)
d\bar N^*(s)
=\dfrac{1}{\sqrt{k_n}}\sum_{i=1}^j {\V}_{\b_0} \left(Z\vert t_i\right)^{-1/2} \mathbf{r}_{\b_0}(t_i).
\end{equation*}
\end{definition}
%%%%%%%%%%%%%%%%%%%%%%%%%%%%%%%%%%%%%%%%%%%%%%%%%%%%%%%%%
 The $\U^*$ process is only defined on $k_n$ equispaced points of the interval
$[0,1]$ but we extend our definition to the whole interval via
linear interpolation so that, for $u$ in the interval $[t_j,t_{j+1}[$, we write 
$$\label{brown2} 
\U^*\left(\b_0,
u\right) = \U^*\left(\b_0, t_j\right) +
\left\{uk_n-j\right\} \left\{\U^*\left(\b_0,
t_{j+1}\right)- \U^*\left(\b_0,t_j\right)
\right\}. 
$$ 
 The following theorem gives the asymptotic behaviour of $\U^*(\b_0,\cdot)$:
 %%%%%%%%%%%%%%%%%%%%%%%%%%%%%%%%%%%%%%%%%%%%%%%%%%%%%%%%%%%%%%%%
\begin{theorem}
\label{thNPH_multi}
Under the non--proportional hazards model of parameter $\b(t)$, we have the following convergence in distribution:
\begin{equation}
\quad \U^*(\b_0,\cdot)-\sqrt{k_n} \C_n\overset{\cal D}{\underset{n\rightarrow \infty}{\longrightarrow}}  \W_p,
\label{derivebrownien_multiv}
\end{equation}
where $\W_p$ is a standard Brownian motion of dimension $p$ and, for all $t \in [0,1]$, 
\begin{equation*}
\C_n(t)=\frac{1}{k_n}\sum_{i=1}^{\lfloor t k_n \rfloor} {\V}_{\b_0} \left(Z\vert t_i\right)^{-1/2}\left\{
{\E}_{\b_0} \left(Z\vert t_i\right)
-
{\E}_{\b(t_i)} \left(Z\vert t_i\right)
\right\}.
\end{equation*}
In addition, we have the convergence of probability 
\begin{equation}
\label{CV_derive_multi}
 \sup_{t \in [0,1]} \left\| \C_n(t)-  \boldsymbol{ \Sigma}^{1/2}\int_0^t\left\{\b(s)-\b_0\right\}ds \right\| \overset{P}{\underset{n\rightarrow \infty}{\longrightarrow}} 0,
  \end{equation}
  where $\int_0^ta(s)ds\!=\!\left(\!\int_0^ta_1(s)ds,\dots,\int_0^ta_p(s)ds\!\right)$ for any $\R^p$--valued function $a\!=\!(a_1,\dots,a_p)$.
\end{theorem}
The proof is given in Appendix \ref{appendix_assproof} and is based on the multivariate functional central limit theorem of \citet{Helland1982}.
The second term of formula (\ref{derivebrownien_multiv}) increases without bound as the sample size goes to infinity. In practical situations, when the model generating the observations is based on $\b(t)$, Theorem \ref{thNPH_multi} in addition to Slutsky's lemma indicate that $\U^*(\b_0,\cdot)$ will look like a multivariate Brownian motion with an added drift term:
\begin{corollary}
\label{corNPH}
Under the model (\ref{nonph}) with parameter $\b(t)$, we have, for all $ \b_0$, 
\begin{equation*}
\U^*(\b_0,\cdot) - \sqrt{k_n}\, \boldsymbol{ \Sigma}^{1/2} IB \overset{\cal D}{\underset{n\rightarrow \infty}{\longrightarrow}} 
 \W_p , 
\end{equation*}
where $IB(t)=\int_0^t\left\{\b(s)-\b_0\right\}ds,\ 0 \leq t \leq 1$.
In addition, $\hat{\boldsymbol{ \Sigma}}=k_n^{-1}\sum_{i=1}^{k_n}\V_{\b_0}(Z\vert t_i)$ is a consistent estimator of $\boldsymbol{ \Sigma}$, and
\begin{equation}
\label{derive_beta}
\hat{ \boldsymbol{ \Sigma}}^{-1/2}\U^*(\beta_0,\cdot) -\sqrt{k_n}IB  \overset{\cal D}{\underset{n\rightarrow \infty}{\longrightarrow}}  \boldsymbol{\Sigma}^{-1/2}\W_p. 
\end{equation}
\end{corollary} 
In the sequel, the standardized score process is evaluated in $\b_0=\bf 0$, where $ \bf 0$ is the null vector of $\R^p$. As a consequence, the plot of $\hat{ \boldsymbol{ \Sigma}}^{-1/2}\U^*({\bf 0},t) $ against the time $t$  gives an indication on the shape of $\int_0^t\b(s)ds$ which is reflected by the shape of the drift of the process (equation (\ref{derive_beta})). In the univariate case ($p=1$) or in the multivariate case with independent covariates, the process $\U^*({\bf 0},\cdot)$ can be directly plotted over time, with no additional standardization since $\Sigma$ is the identity matrix. However, when dealing with correlated covariates, a global standardization is needed for isolating each effect $\beta_i(t)$ on each process $ \left[ \hat{ \boldsymbol{ \Sigma}}^{-1/2}\U^*(\beta_0,\cdot) \right]_i$, $i=1,\dots,p$. A linear drift corresponds to a constant over time regression effect. Our proposed method takes into account the covariances between all covariates and the goodness--of--fit of the overall model is directly evaluated instead of checking proportionality of hazards for one covariate at a time.

\begin{figure}[ht]
\begin{center}
\subfigure[$\beta=0$]{
\includegraphics[width=2.8in,height=2.5in]{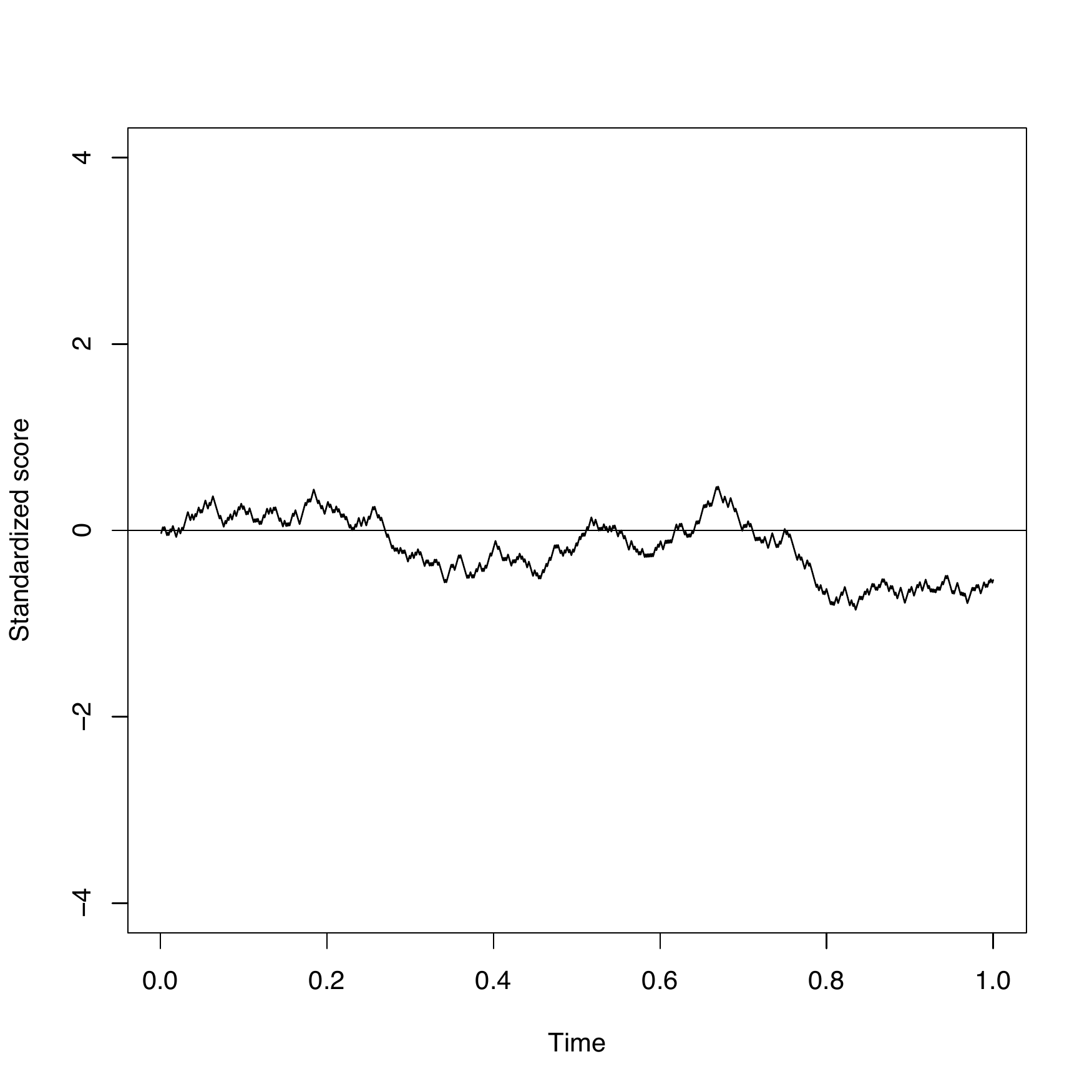}
\label{figure_beta0}}
\subfigure[$\beta=0.5$]{
\includegraphics[width=2.8in,height=2.5in]{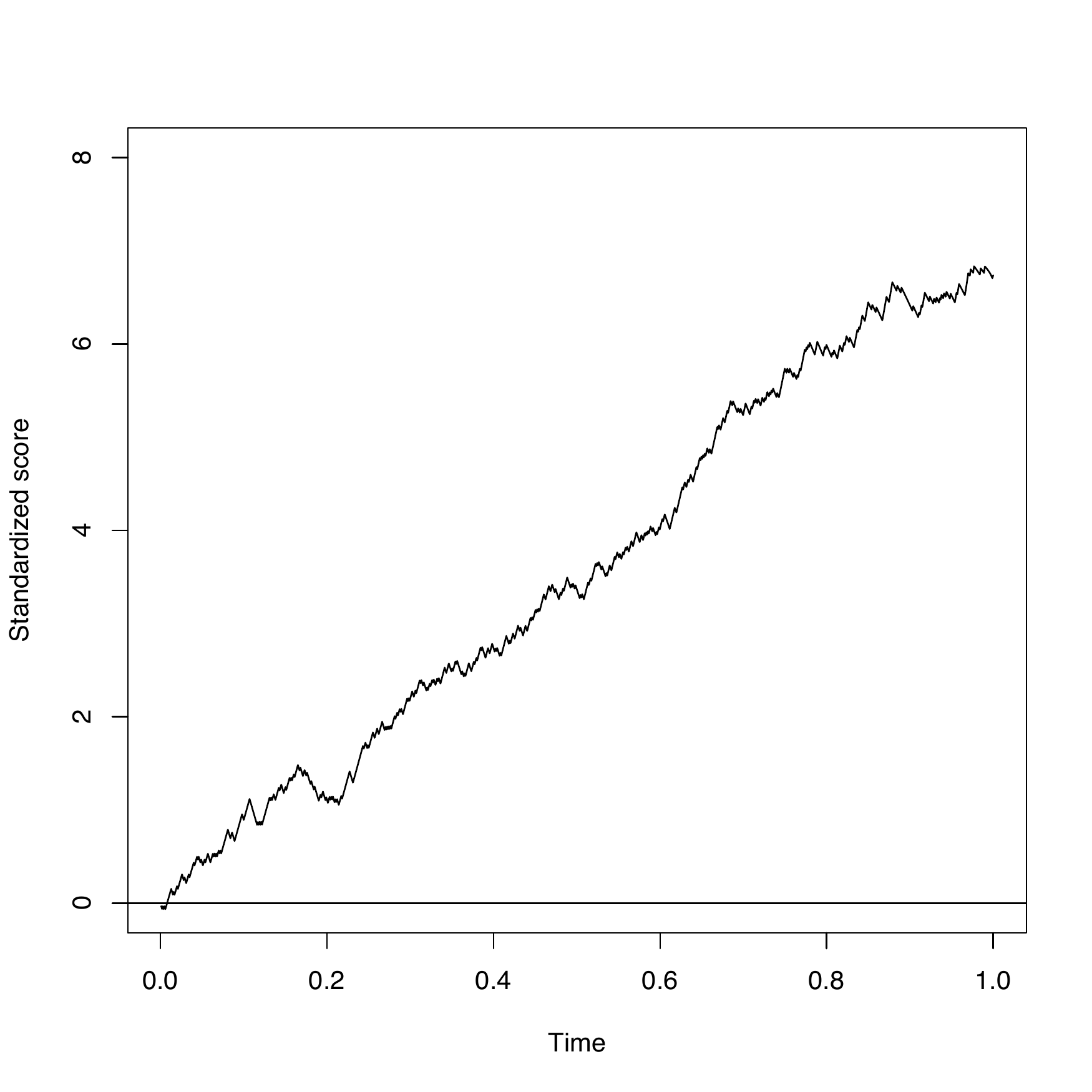}
\label{figure_beta0.5}} 
\caption{Processes $U^*(0,t)$ based on data simulated from proportional hazards models of parameter $\beta$.}
\end{center}
\end{figure}

Illustrations are given in the univariate case. Figure \ref{figure_beta0} represents a simulation of the process $U^*(0,t)$ over time $t$, under the model with a null regression parameter $\beta(t)=0$. Even under moderate to small sample size, the Brownian motion approximation appears accurate enough for reliable inference.
\begin{figure}[ht]
\begin{center}
\subfigure[$\beta(t)=I(t \leq 0.5)$]{
\includegraphics[width=2.8in,height=2.5in]{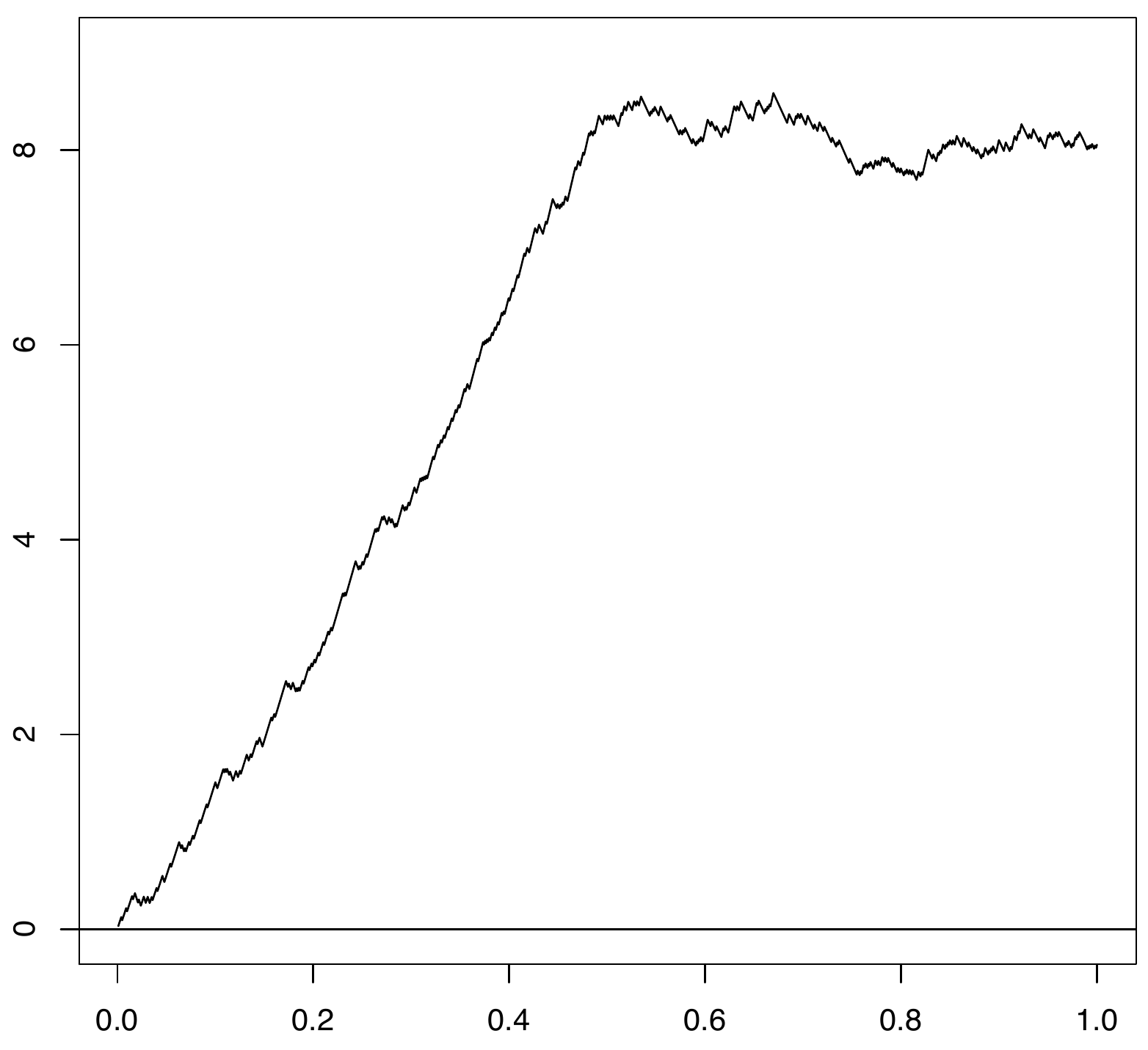}
\label{figure_beta_1_0}}
\subfigure[$\beta(t)=I(t \leq 1/3)+0.5I(t>2/3)$]{
\includegraphics[width=2.8in,height=2.5in]{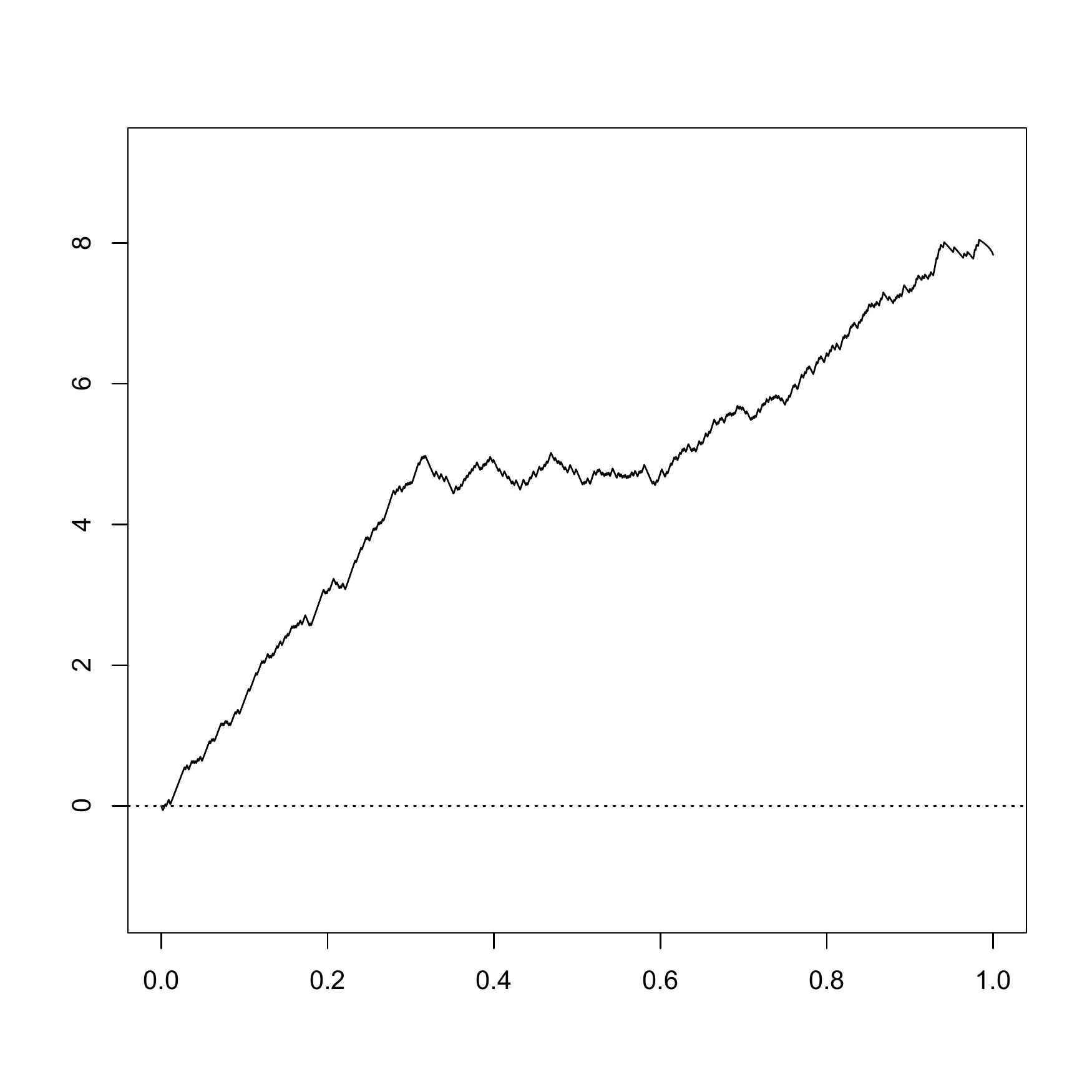}
\label{figure_beta_1_0_05}}
\caption{Processes $U^*(0,t)$ based on data simulated from models (\ref{nonph}) of parameter $\beta(t)$.}
\end{center}
\end{figure}
Consider a proportional hazards model with $\beta(t)$ constant over time but not null. Corollary \ref{corNPH} suggests that a
good approximation for this process is a Brownian motion with a linear drift. An indication of the plausibility of this is shown in Figure \ref{figure_beta0.5}, where $\beta(t)$ is set to $0.5$.
Departures from the proportional hazards assumption can be of various forms. For instance, the effect can be constant and then decreasing after some time $\tau$, the effect can be piecewise constant over time or it can increase over time.
Corollary \ref{corNPH} implies that the shape of the drift of the process $U^*(0,\cdot)$ will reflect the shape of the cumulated regression coefficient.
 As an illustration, Figure \ref{figure_beta_1_0} represents a simulated process under the non--proportional hazards model, with $\beta(t)$ piecewise constant. Before $t=0.5$, there is a linear trend corresponding to $\beta=1$ and for $t>0.5$, $\beta$ equals zero and the process $U^*(0,t)$ is constant in expectation over time. Figure \ref{figure_beta_1_0_05} represents a simulated standardized score process for a changepoint model with the regression parameter $\beta(t)=I(t \leq 1/3)+0.5I(t>2/3) $. The trend of the process can be separated into 3 straight lines reflecting the strength of the effect: the slope of the first part seems twice higher than the one of the last part and the slope of the second part is null. 
The following proposition enables the construction of a confidence band for each process:

\begin{proposition} \label{confband}
Let $i=1,\dots,p$. Consider the hypothesis $H_{0,i}:\exists\, b_i,\,\beta_i(t)=b_i$ and its alternative $H_{1,i}:\nexists\, b_i,\,\beta_i(t)=b_i $.
Under the model (\ref{nonph}) of parameter $\b(t)=(\beta_1(t),\dots,\beta_p(t))$ not necessarily equals to $\b_0$ and $H_{0,i}$, we have, for all $a\geq 0$,
\begin{align} \label{CB_proc}
\lim_{n\rightarrow + \infty} &\text{P}\left( \left\| \hat{\Sigma}_{\cdot,i}^{-1/2} \right\|_2^{-1}\sup_{t\in[0,1]} \left\vert \left(\hat{ \boldsymbol{ \Sigma}}^{-1/2}\left\{\U^*(\b_0,t) -t\U^*(\b_0,1)\right\}\right)_i \right\vert \leq a\right) \nonumber \\
&=\text{P}\left( \sup_{t\in[0,1]} \left\vert B(t)\right\vert \leq a\right),  
\end{align}
where $B$ is a Brownian bridge and $\left\| \hat{\Sigma}_{\cdot,i}^{-1/2} \right\|_2=\left(\sum_{j=1}^p\left(\hat{\Sigma}_{j,i}^{-1/2}\right)^2 \right)^{1/2}$. Therefore, by denoting $a(\alpha)$ the quantile of order $\alpha$ of the Kolmogorov distribution, we have
\begin{equation*}
\lim_{n\rightarrow + \infty }\text{P}\left(\forall t \in[0,1],\ \left[\hat{\Sigma}^{-1/2}\U^*(\beta_0,t)\right]_i \in IC_i(\alpha)\right) =1- \alpha,
\end{equation*}
with 
\begin{equation*} 
IC_i(\alpha)=\left[t\left[\hat{\Sigma}^{-1/2}\U^*(\b_0,1)\right]_i - \left\| \hat{\Sigma}_{\cdot,i}^{-1/2} \right\|_2 a(\alpha) ;t\left[\hat{\Sigma}^{-1/2}\U^*(\b_0,1)\right]_i  -\left\| \hat{\Sigma}_{\cdot,i}^{-1/2} \right\|_2a(\alpha)\right].
\end{equation*}

\end{proposition}

The proof can be found in Appendix \ref{appendix_confband}. If the $i$th element of the process $ \hat{\Sigma}^{-1/2}\U^*(\beta_0,t)$ leaves the confidence band $IC(\alpha)_i$, we reject the hypothesis that the effect $\beta_i(t)$ is constant over time with an asymptotic level of $\alpha$. However, when testing simultaneously several hypotheses $H_{0,i}$ of constant effects for different covariates, the global type I error is inflated. This means that one process could leave its confidence band whereas the corresponding effect is constant over time with a level higher than $\alpha$. This does not seem to be a problem since the plot of the confidence interval is just one of the tools we use to select the variables respecting the proportional hazards assumption. We do not base a definitive conclusion regarding this assumption on this confidence band only, and the non--detection of a constant effect will be corrected with the other steps of the selection variable method we propose in this article.
Of course, corrections for multiple testings could be applied.

Whether effects are of a proportional hazards or a non--proportional hazards form, essentially, all of the information concerning the regression effect $\b(t)$ is captured in the process $\U^*(\mathbf{0},\cdot)$. The process allows the data to speak for themselves, not unlike a scatterplot in linear regression, in which trends and non-linearity may be apparent, since we evaluate the process at $\b_0= \bf 0$. No parameter has to be estimated and expectations and variance-covariance matrices are the usual sequential empirical quantities.
The process, based on the residuals of the non--proportional hazards model, is a useful tool in the evaluation of its goodness--of--fit. These residuals can also be used in the construction of a predictive accuracy measure of the model.
  
\FloatBarrier

\section{INTERPLAY OF FIT AND PREDICTION}

\subsection{$R^2$ Coefficient as a Measure of Predictive Ability}

For any random variables $X$ and $Y$ having second moments, the formula 
\begin{equation}
\label{VarDecomp}
\text{Var}(Y)=E(\text{Var}(Y\vert X))+\text{Var}(E(Y \vert X)),
\end{equation} 
leads to the natural definition of explained variation as the ratio of the variance of the expected values of the response variable under the model given the explanatory variables to the marginal variance of the response variable. In light of the Chebyshev inequality, we see that explained variation directly quantifies predictive strength. 

In the non--proportional hazards model (\ref{nonph}) with one covariate $Z(t)$ ($p=1$), the explained variation makes use of the variance decomposition given in equation (\ref{VarDecomp}) in which $Y$ is replaced by $Z(t)$ and $X$ by $T$, leading to the definition:
\begin{definition} In the univariate non--proportional hazards model (\ref{nonph}), the explained variation, expressed as a function of the time-dependent regression coefficient $\beta(t)$, is defined by
\begin{equation*}
\Omega^2\left(\beta(t)\right)=\dfrac{{\rm Var}(E(Z|T))}{\rm{Var}(Z)} =1-\dfrac{E(\rm{Var}(Z|T))}{\rm{Var}(Z)}.
%\label{EV1}
\end{equation*}
\end{definition}

In the multivariate non--proportional hazards model (\ref{nonph}), individual $i$ is characterized by its real-valued prognostic index $\eta_i(t)=\b(t)^T\Z_i(t)$, being a realization of $\eta(t)=\b(t)^T\Z(t)$. Therefore it is equivalent to evaluate the quality of prediction of the model via $\Z$ or $\eta$. We adopt the latter possibility. 
\begin{definition} The explained variation of the non--proportional hazards model (\ref{nonph}) with multiple covariates can be  defined by a function of the time-dependent regression coefficient $\b(t)$ by
\begin{equation*}
\Omega^2\left(\eta(t)\right)=\dfrac{{\rm Var}(E(\eta|T))}{\rm{Var}(\eta)} =1-\dfrac{E(\rm{Var}(\eta|T))}{\rm{Var}(\eta)}, \quad \eta(t)=\b(t)^T\Z(t).
\end{equation*}
\end{definition}

Some properties of $ \Omega^2$ can be found in \citet[chap. 13]{OQuigley2008} or in \citet[chap. 27]{OQuigley2012}. In these book chapters, $\Omega^2$ is a function of a constant regression parameter $\b$, corresponding to the proportional hazards model. Extension to a time-dependent regression parameter is straightforward. The explained variation coefficient remains constant when applying a monotonically increasing transformation on time. Thus, we work on the standardized time scale as described in Section \ref{sec:process}. \\

\noindent The explained variation is a population parameter that needs to be estimated. Several estimators have been proposed in the literature (\citealt{Choodari2012}). Our goal here is not to present an exhaustive review of these estimators. We focus on the $R^2$ coefficient introduced by \citet{OQuigley1994} since it is built with the same residuals as the standardized score process. We recall its definition by extending it to the non--proportional hazards case.
Let us define the expectation over time of the expected squared discrepancy between the covariate or prognostic index evaluated with parameter $\a_2(t)$ and their expected value under the non--proportional hazards model (\ref{nonph}) of parameter $\boldsymbol \a_1(t)$, not necessarily equals to $\a_2(t)$, of dimension $p$

\begin{align*} 
 Q\big(F&,\a_1 (t),\a_2(t) \big)\\
 &= \left\lbrace
\begin{array}{ll}
\displaystyle{\int_0^1} {E}_{\a_1(t)}\left(\left.\left\{\a_2(t) ^T{Z}(t)-
{E}_{\a_1(t)}\left( \a_2(t) ^T{Z} \mid T=t\right) 
\right\}^2\right\vert T=t\right)dF(t) & \text{if } p>1 \nonumber\\~\\
\displaystyle{\int_0^1} \ {E}_{\a_1(t)}\left(\left.\left\{{Z}(t)-
{E}_{\a_1(t)}\left({Z} \mid T=t\right) \right\}^2\right\vert T=t\right)dF(t) & \text{if }  p=1,
\end{array}
\right.
\end{align*}
where $F$ is the cumulative distribution function of $T$. Then $\Omega^2(\b(t))$ can be expressed as
\begin{equation}
\Omega^2(\b(t))=1-\dfrac{Q(F,\b(t),\b(t))}{Q(F,{\bf 0},\b(t))}.
\end{equation}
Let us denote $\hat F$ the estimator of the cumulative distribution function of $T$ such that $\hat F(t)={k_n}^{-1} \bar N^*(t)$. $ \hat F$ corresponds to the usual empirical cumulative distribution function of $T$ in the uncensored case. Then, $Q\left(F,\a_1(t),\a_2(t)\right)$
can be estimated by
\begin{align*}
\hat{Q}(\hat{F},&\a_1(t),\a_2(t))\\
&=\left\{
\begin{array}{ll}
{\displaystyle \int_0^1} \left\{\a_2(s)^Tr_{\a_1(s)}(s)\right\}^2d\hat{F}(s)=
\dfrac{1}{k_n}\,\overset{k_n}{\underset{i=1}{\sum}}\left\{\a_2(t_i)^Tr_{\a_1(t_i)}(t_i)\right\}^2& \text{ if } p>1 \\~\\
{\displaystyle \int_0^1} \left\{r_{\a_1(s)}(s)\right\}^2d\hat{F}(s)=\dfrac{1}{k_n}\,\overset{k_n}{\underset{i=1}{\sum}}\left\{r_{\a_1(t_i)}(t_i)\right\}^2& \text{ if } p=1. 
\end{array}
\right.
\end{align*}

 The $R^2$ coefficient can then be defined by $R^2=R^2(\hat \b(t))$, where, for all vector $\a(t)$ of dimension $p$,
\begin{equation}
\label{R2def}
R^2(\a(t))=1-\dfrac{\hat Q( \hat F,\a(t),\a(t)) }{\hat Q( \hat F,{\bf 0},\a(t)) }
= \left\lbrace
\begin{array}{ll}
1-\dfrac{\overset{k_n}{\underset{i=1}{\sum}}\left\{\a(t_i)^Tr_{\a(t_i)}(t_i)\right\}^2}{\overset{k_n}{\underset{i=1}{\sum}} \left\{\a(t_i)^Tr_{\bf 0}(t_i)\right\}^2}
& \text{ if } p>1 \\
1-\dfrac{ \overset{k_n}{\underset{i=1}{\sum}} r_{\a(t_i)}(t_i)
^2 }
{\overset{k_n}{\underset{i=1}{\sum}} r_{\bf 0}(t_i)^2 } & \text{ if } p=1.
\end{array}
\right.
\end{equation}

The explained variation coefficient $\Omega^2(\b(t))$ can be estimated by $R^2(\hat{\b}(t)),$ where $\hat{\b}(t)$ is a consistent estimator of the true value of the regression coefficient $\b(t)$. The following theorem will be useful to evaluate the goodness of fit of the model (\ref{nonph}).
\begin{theorem}
\label{theoremR2}
Under the non--proportional hazards model (\ref{nonph}) of parameter $\b(t)$, 
 we have the following convergence
  $$\vert R^2(\b(t))-R^2(\hat \b(t)) \vert\overset{a.s.}{\underset{n\rightarrow + \infty}{\longrightarrow}} 0,$$
and, if $p=1$, with probability one,
$$\underset{{b(t)}}{\arg\max}\ \lim_{n\rightarrow + \infty} R^2(b(t))=\b(t).$$ 
\end{theorem}
The proof is given in appendix \ref{appendix_R2}. The theorem states that if $\b(t)$ is the true regression coefficient, the maximum of $R^2$ is well approximated by $R^2(\hat \b(t))$ for a large enough sample size. This result has been shown for one covariate in the model, and in the case of multiple covariates, we conjecture an analogous result for the one--dimensional prognostic index.
 The predictive ability measure and the standardized score process are built with the same ingredients.  
 The standardized score process enables to check the fit of the model, whereas the $R^2$ coefficient is a measure of the predictive ability of the %non--proportional 
 model. Although different, these two aspects of the model are related; their construction with the same quantities seems then to be quite natural. 

\subsection{Using the $R^2$ Coefficient to Improve the Fit}
Using the results of Theorem \ref{thNPH_multi} and its corollary, the standardized score process can be used to determine the shape of the temporal regression effect. No other tools such as smoothing, the projection on a basis of functions or kernel estimation are needed
\citep{Cai2003,Hastie1990,Scheike2004}. For instance,  as shown in Figure \ref{figure_beta_1_0}, a constant effect until time $\tau$ followed by a null effect is easily detectable, especially with moderate and larger sample sizes. Assume that the time-dependent regression parameter can be expressed as $\b(t)=(\beta_1(t),\dots,\beta_p(t))$, where $\beta_j(t)=\beta_{0,j} \, B_j(t)$ ($j=1,\dots,p$) with $\b_0=(\beta_{0,1},\dots,\beta_{0,p})\in \R^p$ an unknown regression parameter and $\mathbf{B}=(B_1,\dots,B_p)$ a known $\R^p$-valued function of the time. Thus, $\widehat{ \b}(t)= (\widehat{ \beta_1}(t),\dots,\widehat{ \beta_p}(t))$ where $\widehat{ \beta_j}(t)=\widehat{ \beta_{0,j}}\, B_j(t)$ ($j=1,\dots,p$) with $\widehat{ \b_0}$ the maximum likelihood estimator of $\b_0$, obtained via classical maximization of the partial likelihood \citep{Cox1972}. The function $B(t)$ can determined graphically using the standardized score process (see the examples of Section \ref{section_applications}). In addition, the confidence bands defined in Proposition \ref{confband} can help to evaluate the plausibility of a constant effect $\beta_j(t)$ over time resulting in a constant function $B_j$, for each $j=1,\dots,p$.

When dealing with non--proportional hazards, the investigator needs an instrument other than one that is focused solely on fit. This can be provided by the $R^2$ coefficient that not only indicates predictive strength but will tend to a maximum value when the correct form of $B(t)$ is chosen (Theorem \ref{theoremR2}). 
When different competing models provide plausible forms for $B(t)$, the one maximizing the $R^2$ coefficient would be considered the best. Using this procedure, we obtain a non--proportional hazards model with a good fit and a maximal predictive ability. 
The predictive ability measure is maximized on the set $\cal B$ of the temporal regression effects selected by the investigator. Formally,
let $\mathcal{B}=\{\b_1(t),\dots,\b_m(t)\}$ be a set of
$m$ functions from $[0,1]$ to $\R^p$.
The selected regression function $\b^*(t)$ is such that
$$\b^*(t)=\arg\max_{b(t) \in \mathcal{B}}R^2\left(b(t)\right)  .$$
The following theorem gives an equivalence between this maximization problem when $n \rightarrow \infty$ and a problem of minimization of $L^2$ norms.

\begin{theorem} 
\label{theoremR2L2}
Let $p=1$. Under the non--proportional hazards model (\ref{nonph}) with regression parameter $\beta(t)$ not necessarily in $\cal B$, asymptotically, $ \beta^*(t)=\underset{\alpha(t) \in \mathcal{B}}{\arg\max}\ \underset{n\rightarrow \infty}{\lim}R^2\left(\alpha(t)\right)$ is the solution of
$$ \beta^*(t)=\arg\min_{\alpha(t) \in \mathcal{B}} \left\| \beta(t)-\alpha(t)\right\|_{2,W},$$
where $  \left\| a(t)\right\|_{2,W}=\left(\int_0^1a(t)^2v(c(t),t)^2dt\right)^{1/2}$ is a weighted $L^2$ norm of the function $a(t)$ from $[0,1]$ to $\R$, with $c(t)$ lying between $ 0$ and $a(t)$.\newline
\end{theorem}

%%%%%%%%%%%%%%%%%%%%%%%%%%%%%%%%%%%%%%%%%%%%%%%%%%%%%%%%%%%%%%%%%%%%%%%%
Proof can be found in Appendix \ref{proofR2L2}. In other words, for large enough sample sizes, selecting the regression coefficient by maximizing the $R^2$ coefficient is the same as selecting the closest temporal regression function to the true coefficient in the $L^2$ norm sense. 

 A model is chosen to fit a dataset because of either a good fit or a good predictive capacity. Several models could present one of these aspects or both of them, not only the "true" model. 
 Priority is given to the goodness of fit, with the selection of possible time-dependent coefficients, and in a second phase, the predictive capacity is considered. 
 We have chosen to work with the $R^2$ coefficient but notice that other predictive ability measures verifying Theorem \ref{theoremR2} might be considered. 
When the trend of the process is a concave function, the effect disminishes over time, whereas in presence of a convex function, the effect increases. In order to obtain the largest possible $R^2$, we could create a temporal effect matching more and more closely the observed trend of the process, e.g. piecewise constant effects with multiple changepoints. In general, this would result in an overfit. In this case, the interpretation of the coefficient is not clear. A tradeoff has to be established between a high predictive ability and the simplicity of the coefficient, especially regarding its interpretation.  This parallels linear regression where the estimated explained variation is positively biased %for the true unknown explained variation 
and this bias increases with the dimension of the model. Some balance needs to be struck between the goal of improved prediction and the dangers of over optimistic predictions as a result of over fitting.

\section{SOME SIMULATED EXAMPLES}
\label{section_applications}
The simulations are performed with a moderate sample size set to $n=200$ subjects and $\lambda_0(t)=1$. All cases presented here are uncensored. The effect of an independent censoring mechanism on the process is the same as a reduction in the sample size. 
\subsection{Univariate cases}
In both considered cases, the covariate follows a Bernoulli distribution of parameter $0.5$. 
First, we consider the proportional hazards situation by setting $\beta(t)=1.5$.
\begin{figure}[h]
\label{fig_PH} 
\center
\includegraphics[width=2.5in,height=2.5in]{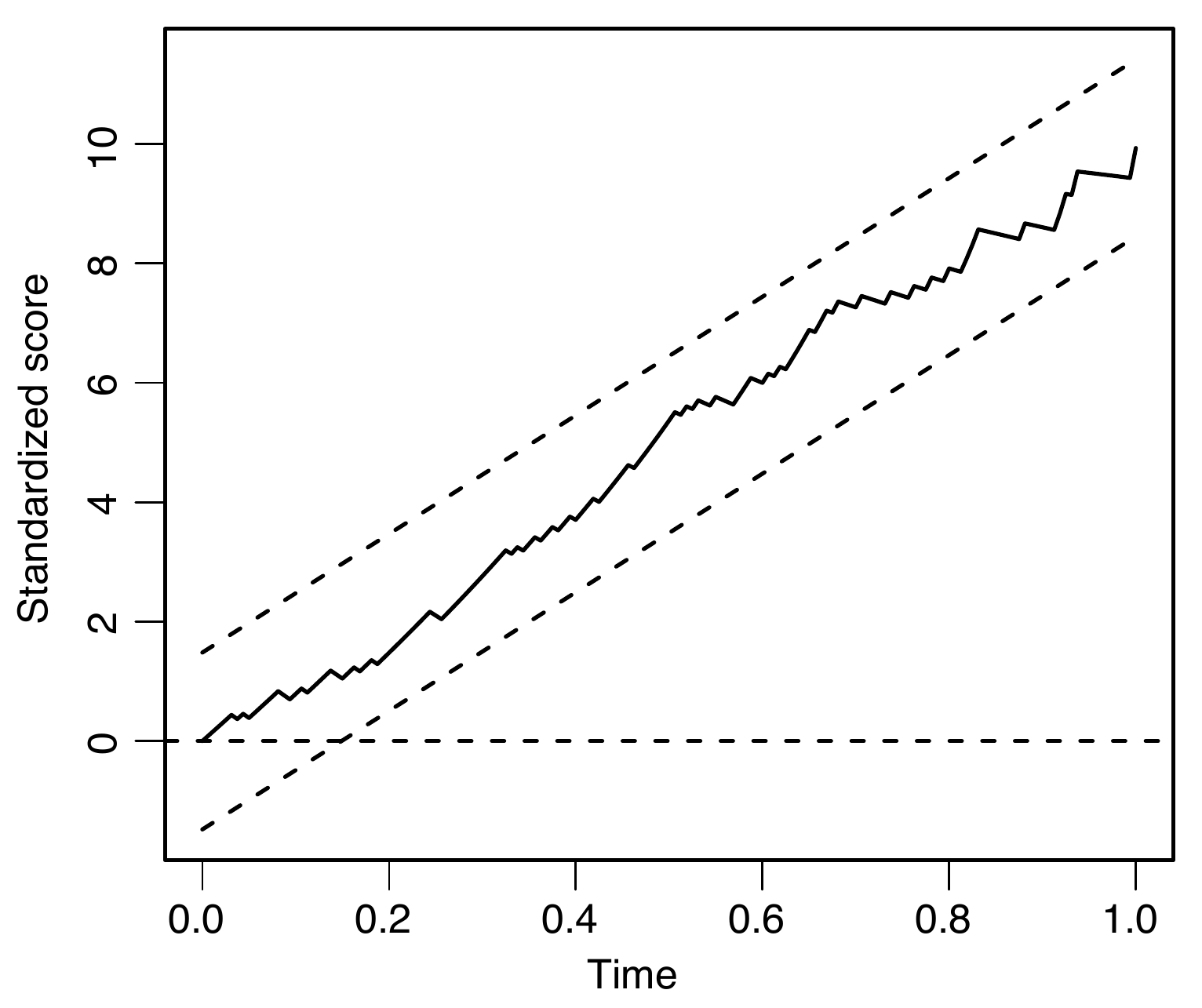}
\caption{Standardized score process $U^*(0,\cdot)$ (solid line) and confidence bands (dotted lines) 
on a simulated dataset with constant regression coefficient $\beta(t)=1.5$.}
\end{figure}
 The standardized score process $U^*(0,\cdot)$ (solid line) and its confidence bands under proportional hazards assumption (dotted lines) are plotted over time in Figure \thefigure. A drift is observed, the effect is not null. The drift seems linear and the process stays between the confidence bands: 
the hypothesis of a proportional hazards model seems reasonable. The usual maximum partial likelihood estimator is estimated at $1.59$ which gives an $R^2$ of $0.35$.

\FloatBarrier

\begin{figure}[h]
\label{fig_3(1-t)2_a}
\center
\includegraphics[width=2.5in,height=2.5in]{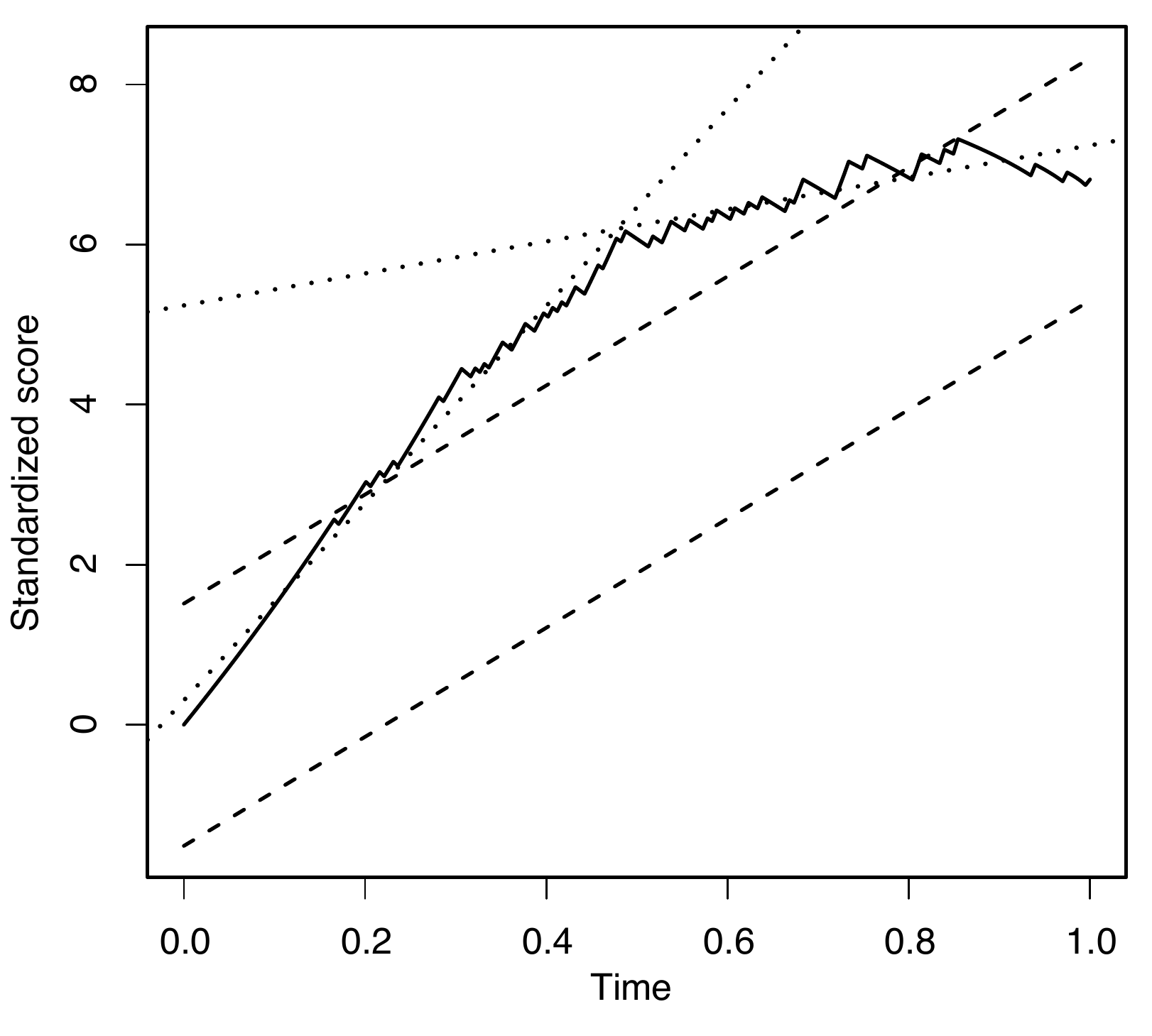} 

\caption{
Standardized score process $U^*(0,\cdot)$ (solid line), confidence bands (dashed lines) 
and a fitted changepoint model (dotted lines) on a simulated dataset with  $\beta(t)=3(1-t)^2$.}
\end{figure}

The next case deals with a smooth decreasing effect. We simulate a dataset with $\beta(t)=3(1-t)^2$. The resulting standardized score process $U^*(0,\cdot)$ (solid line) is plotted over time in Figure \thefigure $ $ with its confidence bands under proportional hazards assumption (dotted lines). The process leaves the confidence bands which indicates that the proportional hazards assumption does not hold. The concavity of the trend gives an indication regarding the decrease of the effect. Amongst other possibilities, the effect could be linear, of a quadratic shape or a piecewise constant function of the time. In the latter case, the trend appears linear up to time $t=0.5$ corresponding to a constant coefficient. Then, the drift changes to a lower constant value, corresponding to a coefficient $\tilde \beta(t)=\beta_0\{I(t\leq 0.5)+C.I(t>0.5)\},$ where $\beta_0$ and $C$ are unknown. $C$ is the value by which the coefficient is multiplied in the second part of the study. In Figure \thefigure
, using linear regression, two straight dotted lines have been fitted to the process, before and after the changepoint time $t=0.5$. The ratio of the second slope over the first one is the value $C=0.16$. Various models with decreasing effect $\beta(t)=\beta_0 B(t)$ have been selected, their $R^2$ coefficients  and $\hat \beta_0$ the maximum partial likelihood estimator of $\beta_0$ have been evaluated in Table \ref{table_3(1-t)2}. The lowest $R^2$ coefficient corresponds to the proportional hazards model and the largest $R^2$ coefficient of Table \ref{table_3(1-t)2} is the one associated with the model of regression coefficient $\beta(t)=\beta_0(1-t)^2$, with an estimation of $\beta_0$ equals to $0.37$. Using our procedure, the regression coefficient used to create the dataset has been selected.

\begin{table}[h]
\center
\begin{tabularx}{15cm}{|X||X|X|X|X|X|}
\hline $\beta(t)$ & $\beta_0$ & $\beta_0(1-t)$ & $\beta_0(1-t)^2$ & $\beta_0(1-t^2)$ & $\tilde \beta(t)$  \rule[-7pt]{0pt}{12pt} \\
\hline
\hline $\hat{\beta_0}$ &  1.06 & 2.45 & 3.73 & 1.77 & 1.83  \rule[-7pt]{0pt}{12pt} \\
\hline $R^2$ & 0.25 & 0.36 & 0.37 & 0.34 & 0.34 \rule[-7pt]{0pt}{12pt}\\ 
\hline
\end{tabularx}
\caption{
Maximum partial likelihood estimators $\hat \beta(t)$ and $R^2$ coefficients
on a simulated dataset with $\beta(t)=3(1-t)^2$.}
\label{table_3(1-t)2}
\end{table}

\subsection{Multivariate case}
\begin{figure}[h]
\center
\subfigure[Regression effect $\beta_1(t)$]{
\includegraphics[width=2.5in,height=2.5in]{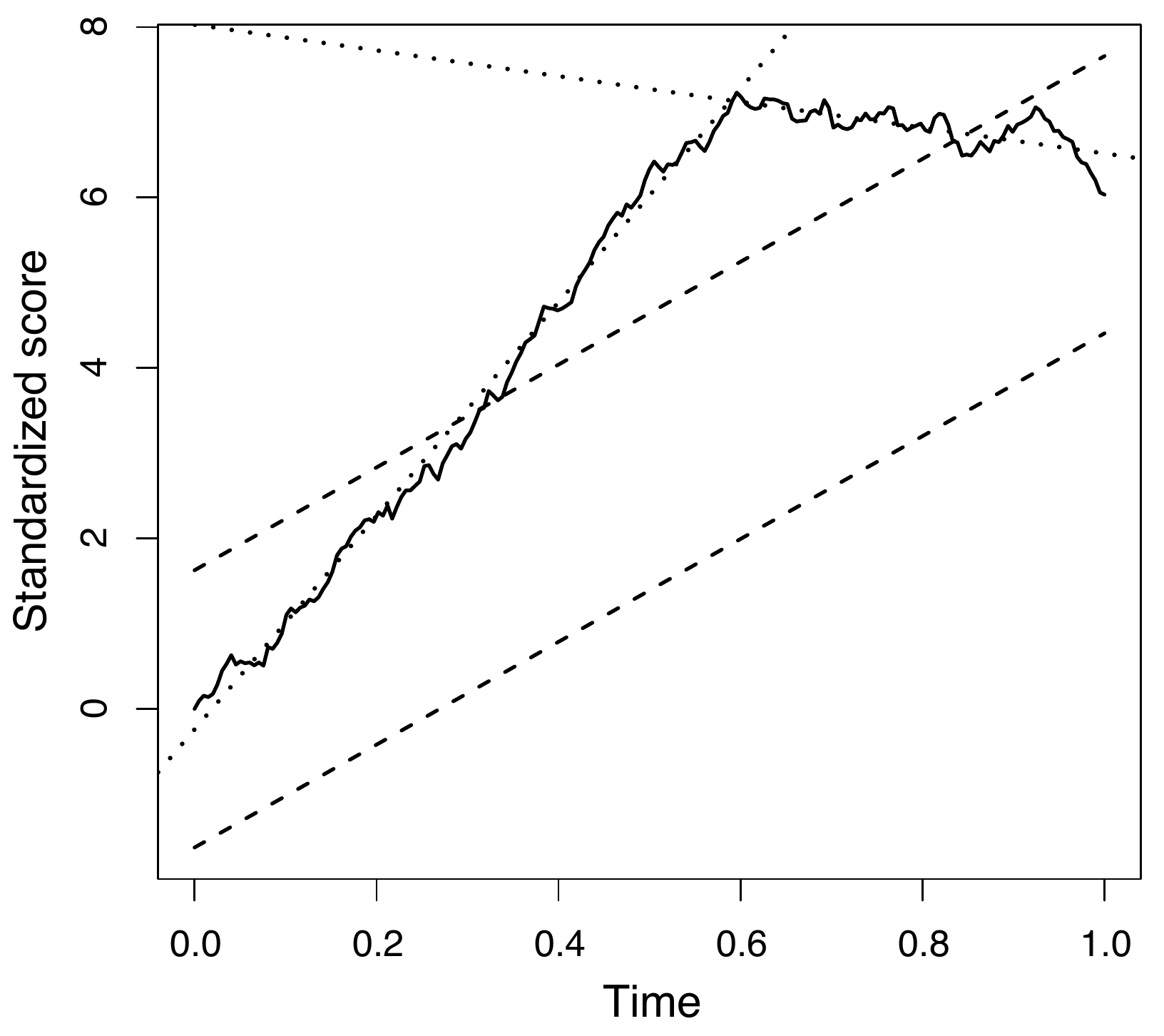}
\label{fig_multiv_a}
}
\subfigure[Regression effect $\beta_2(t)$]{
\includegraphics[width=2.5in,height=2.5in]{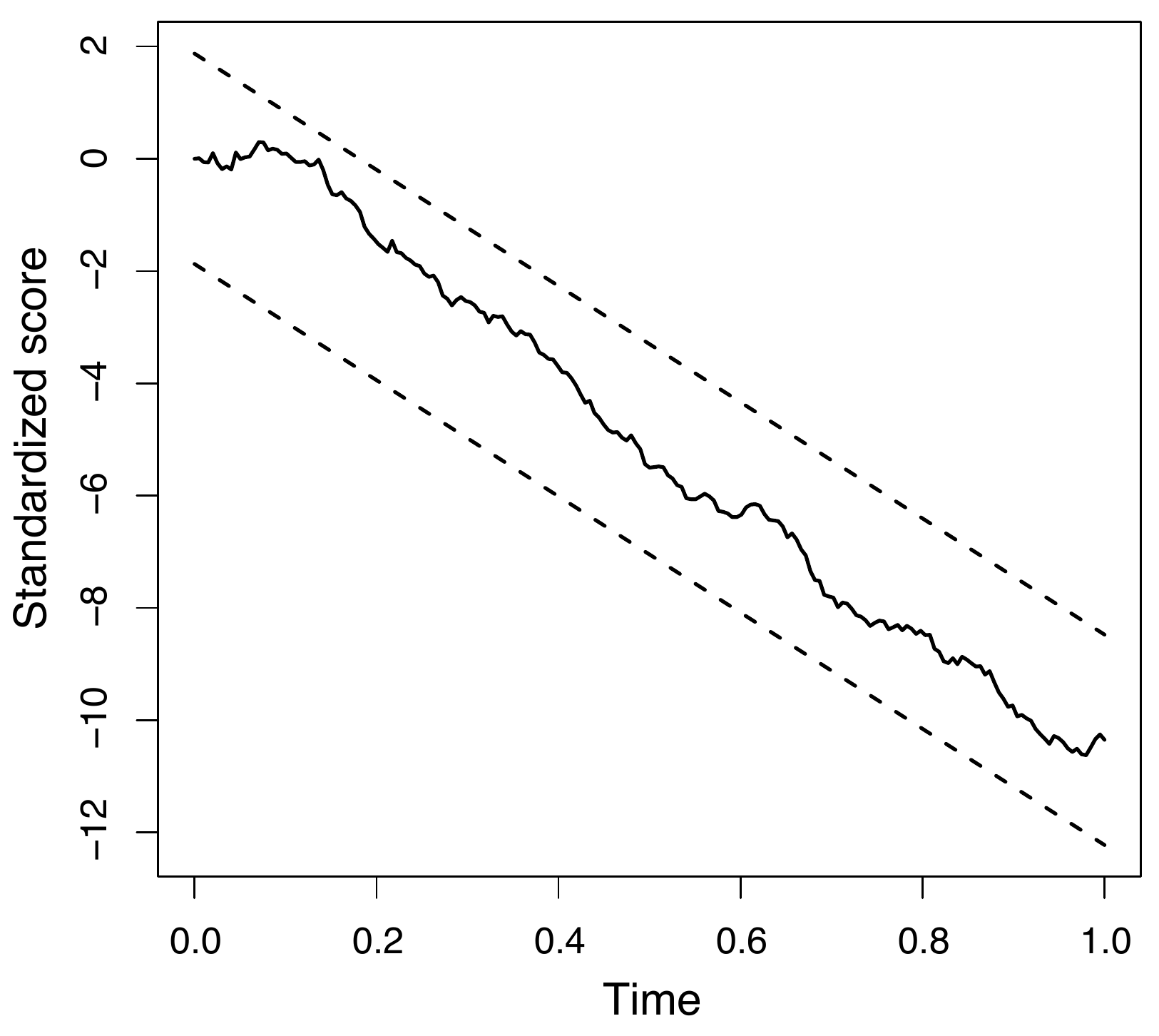}
\label{fig_multiv_b}}
\caption{Standardized score process $\hat \S^{-1} \U^ *(0,\cdot)$ (solid line), confidence bands (dashed lines) 
and a fitted changepoint model (dotted lines) on a simulated dataset with $\beta_1(t)=I(t\leq 0.5)$ and $\beta_2(t)=-1$.
}
\end{figure}

We simulate two standard normal covariates $Z^1$ and $Z^2$ with covariance equals to $0.5$. We set $\beta_1(t)=I(t\leq 0.5)$ and $\beta_2(t)=1$. Each component of the bivariate process $\hat \S^{-1/2}\U^*(\mathbf{0},\cdot)$ (solid lines) is plotted over time on Figures \ref{fig_multiv_a} and \ref{fig_multiv_b} with the confidence bands (dotted lines). 
Clearly, the proportional hazards assumption is rejected for covariate $Z^1$ since the process leaves the confidence band. The shape of the process indicates a piecewise constant regression coefficient, with a changepoint at time $t=0.6$. As in the univariate case, two straight (dashed) lines have been fitted to the process, one before $t=0.6$ and one after. The ratio of the slopes is $-0.12$ which makes us consider the regression coefficient $\beta_1(t)=\beta_1\,B_{0.6}(t)$ where $B_{0.6}(t)=I(t\leq 0.6)-0.12\,I(t\geq 0.6)$. Other piecewise constant regression coefficients $\beta(t)=\beta_1\,B_{t_0}(t)$ have been considered with changepoints at times $t_0\in\{0.45,0.5,\dots,0.7\}$. For each time $t_0$, the ratio of slopes has been evaluated to determine the value which multiplies the coefficient in the second part on the study. The second covariate $Z^2$, however, seems to have a constant regression coefficient since the process stands between the confidence bands and 
has a linear trend (Figure \ref{fig_multiv_b}). Therefore, we consider only the regression coefficient $\beta_2(t)=\beta_2$. 
\begin{table}[h]
\label{table_multiv}
\center
\begin{tabular}{|c||c|c|c|c|c|c|c|c|}
\hline $\beta_1(t)$ & $ \beta_1$ &$ \beta_1\,B_{0.45}(t)$  &$ \beta_1\,B_{0.5}(t)$  &$ \beta_1\,B_{0.55}(t)$  & $ \beta_1\,B_{0.6}(t)$  & $ \beta_1\,B_{0.65}(t)$  &$ \beta_1\,B_{0.7}(t)$ \rule[-7pt]{0pt}{12pt} \\
\hline $\hat{\beta}_1$ & 0.45 & 0.93 & 0.96 & 0.89 & 0.95 & 0.86 & 0.72\rule[-7pt]{0pt}{12pt}\\
\hline $\hat \beta_2$ & -0.73& -0.72& -0.73& -0.74& -0.79& -0.80& -0.77\rule[-7pt]{0pt}{12pt} \\
\hline $R^2$ & 0.24 & 0.35 & 0.37 & 0.35 & 0.39 & 0.37 & 0.32 \rule[-7pt]{0pt}{12pt}\\
\hline
\end{tabular}
\caption{Maximum partial likelihood estimators $\hat \beta(t)$ and $R^2$ coefficients
on a simulated dataset with $\beta_1(t)=I(t\leq 0.5)$ and $\beta_2(t)=-1$.}
\end{table}
Estimation results are given in Table \thetable
. The proportional hazards model gives an $R^2$ of 0.24. The maximal $R^2$ is obtained when considering $\beta_1(t)=\beta_1\,B_{0.6}(t)$, with an increase of $60\%$ compared to the proportional hazards model. Therefore, we choose the model with $\beta_1(t)=\beta_1\,B_{0.6}(t)$ and $\beta_2(t)=\beta_2$.

\section{CLINICAL STUDY IN BREAST CANCER}

\begin{figure}[h]
\label{fig_breastcanc}
\center
\subfigure[Tumor size]{ \label{fig_breastcanctaille}
\includegraphics[width=2.5in,height=2.5in]{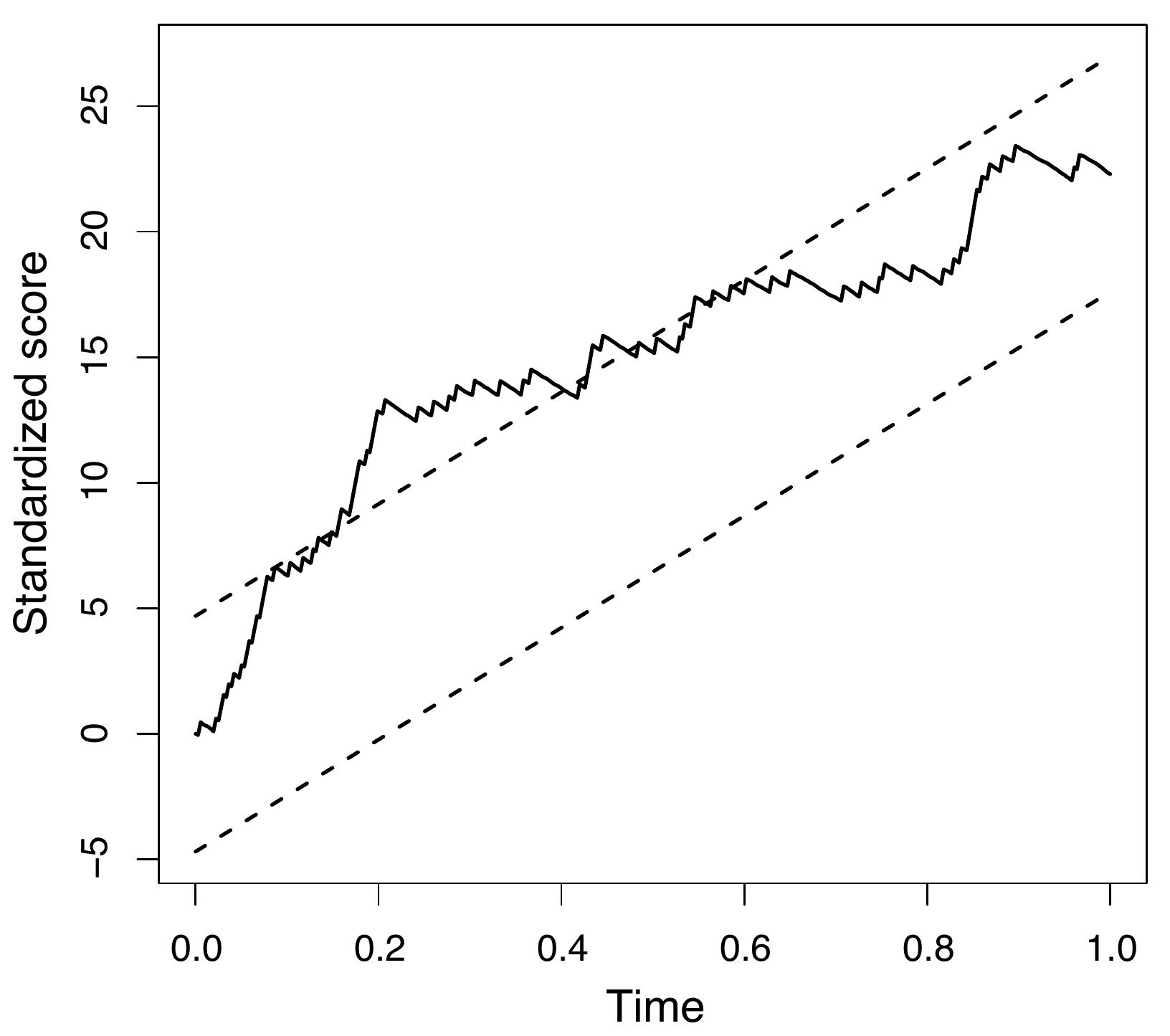}}
\subfigure[Progesterone receptor]{ \label{fig_breastcancrec}
\includegraphics[width=2.5in,height=2.5in]{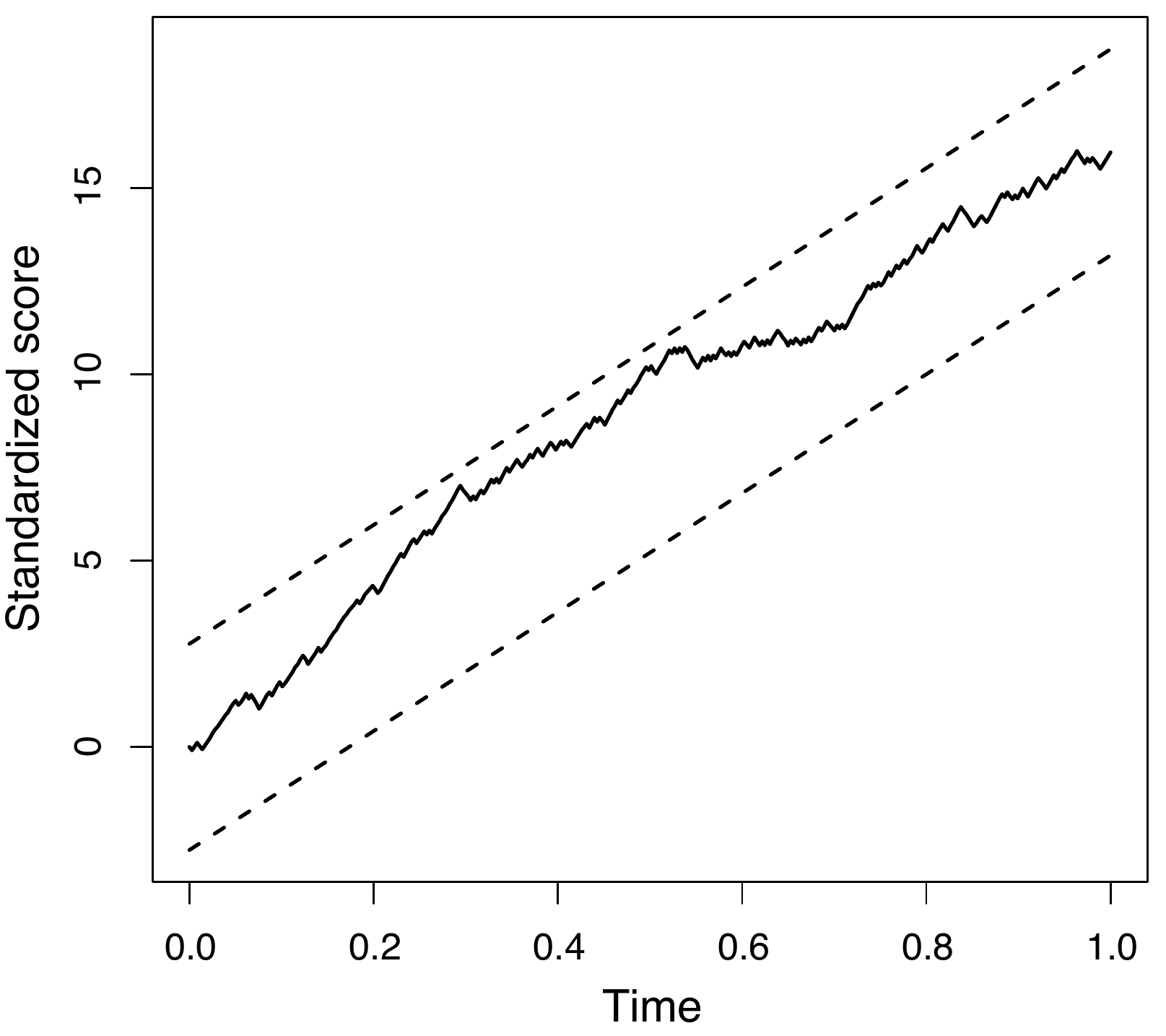}}

\subfigure[Grading]{ \label{fig_breastcangrade}
\includegraphics[width=2.5in,height=2.5in]{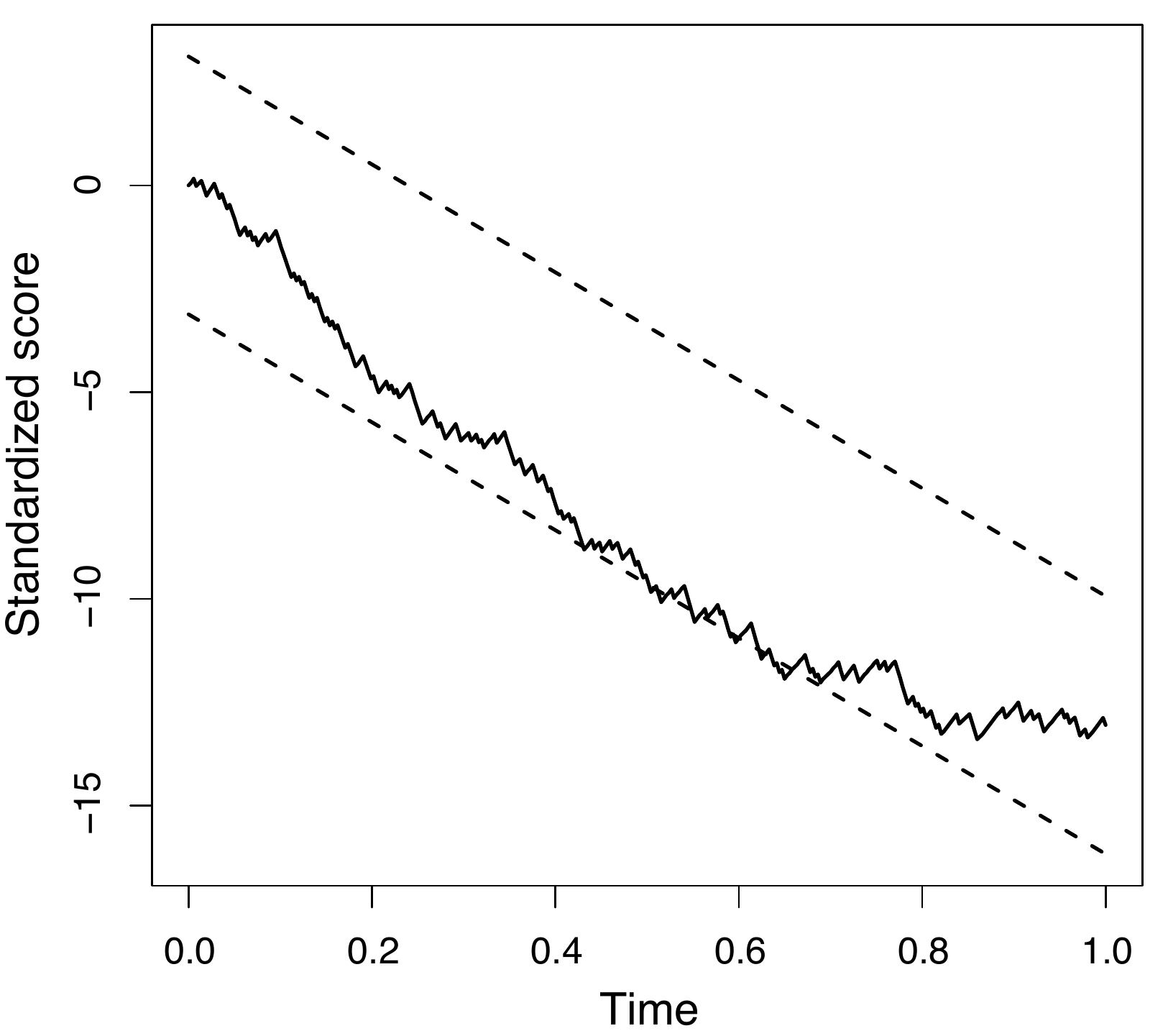}}
%breastcancer.pdf} 
\caption{Standardized score process $\hat \S^{-1} \U^ *(0,\cdot)$ (solid lines) and its confidence bands (dashed lines) on the breastcancer dataset for tumor size, progesterone receptor and grading.}
\end{figure}

We return to the motivating example of the $1504$ patients suffering from breast cancer. These patients were followed over a period of 15 years at the Institut Curie in Paris, France. Several studies were based on these data. One sub-study considered the predictive effects of the prognostic factors; progesterone receptor status, the tumor size over $60$ mm and the grading over 2. The multivariate standardized score process $\hat \S^{-1/2}\U^*(\mathbf{0},\cdot)$ and its confidence band are plotted over time in Figure \thefigure
. In Figure \ref{fig_breastcanctaille}, we illustrate the process corresponding to the tumor size effect. Clearly, the effect seems non--constant with slope gradually diminishing with time. So much so that the process ends up drifting beyond the limits of the 95$\%$ confidence band. A slightly more refined model providing a much better fit allows for a change in effect at time point $t=0.2$. As in our simulated examples, two straight lines have been fitted to the curve before and after $t=0.2$, leading us to consider the regression effect $\beta_{size}(t)=\beta_0(I(t\leq 0.2) +0.24 I(t\geq 0.2))$. From Table 3 %\ref{table_breastcanc} 
we can quantify the predictive improvement of Model 2 (constant effects for hormone receptor status and grade, time dependent effects for tumor size) versus Model 1 (all 3 prognostic factors constant) by a greater than $30\%$ increase in the size of $R^2$, from $0.29$ to $0.39$.  
Figure \ref{fig_breastcancrec} represents the process for the effect of the progesterone receptor over time. Again there is some evidence of a changing slope, although much weaker than for tumor size and, indeed, the process remains within the limits of the confidence bands. We considered various potential regression effects: a changepoint model with a cut at time $t=0.5,$ $\beta_{rec0}(t)=\beta_0(I(t\leq 0.5) +0.39 I(t\geq 0.5))$ and several smooth parameters $\beta_{rec1}(t)=\beta_0(1-t)$, $\beta_{rec2}(t)=\beta_0(1-t)^2$, $\beta_{rec3}(t)=\beta_0(1-t^2)$ and $\beta_{rec4}(t)=\beta_0\log(t)$. 
Figure \ref{fig_breastcangrade} represents the process for the grading effect. There is a clear impression of the steepness of the negative slope attenuating with time. The process reaches the limits of the confidence bands but does not go beyond them. The simpler model, i.e., proportional hazards effects implying a linear slope, may be good enough although, in a model building context, it is also worth considering one with time dependent effects. Specifically, we chose to also look at a model with piecewise constant coefficients $\beta_{gra}(t)=\beta_0(t)(I(t\leq 0.4) +0.69 I(t\geq 0.4))$. 

All of these several combinations, alongside models with constant effects, were looked at. For each combination, the regression effects have been estimated by maximizing the partial likelihood and the $R^2$ coefficient has been evaluated. 
\begin{table}[h]
\label{table_breastcanc}
\center
\begin{tabular}{|c|c|c|c|}
\hline
Tumor size & Receptor & Grading & $R^2$ \\
\hline 0.84 & 1.03 & -0.68 & 0.29 \\
\hline $1.77(I(t\leq 0.2) +0.24 I(t\geq 0.2))$ & 1.03 & -0.66 & 0.39 \\
\hline 0.85 & $-1.02\log(t)$ &  -0.67 & 0.39 \\
\hline $1.74 (I(t\leq 0.2) +0.24 I(t\geq 0.2))$ & $-1.02\log(t)$ & -0.66 &0.51 \\
\hline $1.72 (I(t\leq 0.2) +0.24 I(t\geq 0.2))$ & $-1.02\log(t)$ &   $-0.82I(t\leq 0.4) +0.69 I(t\geq 0.4)$ & 0.52 \\
\hline
\end{tabular}
\caption{Maximum partial likelihood estimators and $R^2$ coefficients
on the breast cancer dataset.}
\end{table}
Partial results are given in Table \thetable
. The proportional hazards model gives an $R^2$ coefficient of $0.29$. As mentioned above, a more involved model allowing for the effect of tumor size to assume a simple time dependency results in a big jump in observed predictability of an order greater than 30\%.The highest $R^2$ is obtained with changepoints for tumor size and grading covariates, with a function of $\log(t)$ for the effect of progesterone receptor. The predictive accuracy of this model has increased by $80\%$ compared to the predictive accuracy of the corresponding proportional hazards model. This gives a strong indication that, as far as prediction is concerned, significant improvement can be consequent on allowing time dependency. On the other hand, allowing for time dependency  grade, having already accounted for the joint effects of tumor size and receptor status, results in an increase in $R^2$ from 0.51 to 0.52. Such a negligible increase dose not justify the added complexity of the model so that, provided the other two risk factors are included, it makes sense to restrict the effects of grade to be constant.

\FloatBarrier

\section{DISCUSSION}
The related and complementary techniques of goodness of fit and predictive ability provide a coherent way to construct models. Intuitively, models constructed in this way ought provide a better predictive performance. This intuition is correct and is supported by the theoretical results of this paper. Our preference is to appeal to techniques based on the Schoenfeld residual processes for proportional and non--proportional hazards models since these processes provide the basis for both of these techniques. A large number of competing approaches appears possible since there is a large body of literature on goodness of fit procedures and a large body on predictive measures. Combinations of these could provide tools analogous to those described here. However, in order to make analogous claims to ours concerning predictive performance for some particular combination, we would require equivalent theorems to those presented in Sections 2 and 3.

We might consider that the first step away from a proportional hazards model is a similar model but with a changepoint. Before the changepoint we have one particular proportional hazards model whereas, after the changepoint, we have a model with a different value of $\beta .$ The methods described here would enable us to estimate the changepoint itself as well as the values of $\beta,$ before and after the changepoint. Extending this to more than a single changepoint is, at least in theory, straightforward. This suggests one possible systematic way of model construction. Another extension that would be worth considering is the estimation of the process drift with non--parametric estimation techniques in order to estimate the cumulative regression effect $\int_0^t\b(s)ds$.

\FloatBarrier

\appendix
\section{Proof of Theorem \ref{thNPH_multi} }
\label{appendix_assproof}

 %%%%%%%%%%%%%%%%%%%%%%%%%%%%%%%%%%%%%%%%%%%%%%%%
Define the filtration $\left\{{\cal F}_t \right\}_{t \in [0,1]}=\sigma\{ \bar N_j(u),\bar Y_j(u^+),\Z_j(u^+),\, j=1,\dots,n, \, 0\leq u\leq t \}.$ Each failure time $t_i$ is a $\mathcal{F}_t$-stopping time. Consider the conditional expectation $$\E_{\b(t)}\left(\mathbf{h}\vert \mathcal{F}_t\right)=\sum_{j=1}^n \mathbf{h}_j(t)\pi_j(\b(t),t), $$ where $\mathbf{h}_j$ is a $\R$ or $\R^p$--predictable process for individual $j$. In order to simplify the notation, denote $\E_{\b(t)}(\mathbf{h}\vert t)=\E_{\b(t)}\left(\mathbf{h}\vert \mathcal{F}_t\right)$ and $\V_{\b(t)}(\mathbf{h}\vert t)=\E_{\b(t)}\left(\mathbf{h}^{\otimes 2}\vert \mathcal{F}_t\right) -\E_{\b(t)}(\mathbf{h}\vert \mathcal{F}_t)^{\otimes 2}. $ 
The first part of the proof shows the convergence in distribution of $\U^*(\beta_0,\cdot)-\sqrt{k_n} \mathbf{C}_n$ to a multivariate Wiener process as $n$ increases without bound. Denote $\X_n$ the right-continuous with left-hand limits process $\X_n$, with a jump at each $t_i$ such that 
\begin{eqnarray*}
\X_n(t)=\dfrac{1}{\sqrt{k_n}}\sum_{i=1}^{\lfloor t k_n \rfloor} \V_{\b_0}\left( Z \vert t_i\right)^{-1/2}\left\{ \mathcal{Z}(t_i)- \E_{\b(t_i)}\left( Z \vert t_i\right) \right\}, \quad 0\leq t\leq 1.
\end{eqnarray*}
Notice that $\X_n(t_i)= \U^*(\b_0,t_i)-\sqrt{k_n} \mathbf{C}_n(t_i)$ at each $t_i=i/{k_n}$, $i=1,\dots,k_n$.
Denote $\bxi_{i,k_n}=(\xi_{i,k_n}^1,\dots,\xi_{i,k_n}^p)$ the $i$th $\R^p$-valued increment of the process $\X_n$. Notice that $\bxi_{i,k_n}$ is ${\mathcal{F}}_{t_i}$-measurable. 
Then,
\begin{align*}
&\left\| \U^*(\b_0,\cdot)-\sqrt{k_n} \mathbf{C}_n -  \mathbf{W}_p \right\|
\leq \left\| \U^*(\b_0,\cdot)-\sqrt{k_n} \mathbf{C}_n - \X_n \right\| +\left\| \X_n - \mathbf{W}_p \right\|.
\end{align*}
The first term on the right hand side converges to 0 as $n$ increases without bound by 
the existence of a moment of order 3 of the increments $\bxi_{i,k_n}$. The convergence in distribution of $\X_n$ to $\W_p$ is given by the multivariate functional central limit theorem of \citet{Helland1982} of which hypotheses are verified in Supplementary Material. It remains to prove equation (\ref{CV_derive_multi}). A multidimensional Taylor-Lagrange series expansion gives
\begin{equation*}
\left\|\E_{\b(t)}(Z\vert t) -\E_{\b_0}(Z\vert t) -\V_{\b_0}(Z\vert t)\left\{\b(t)-\b_0\right\} \right\| \leq \dfrac{M_n}{2} \left\| \b(t)-\b_0 \right\|^2.
\end{equation*}
Therefore,
%%%%%%%%%%%%%%%%%%%%%%%%%%%%%%%%%%%%%%%%%%%%%%%%%%%%%%%%%%%%%%%%%%%%%%%%%%%%%%%%
\begin{align*}
&\left\|C_n(t)-\S^{1/2}\int_0^t\left\{\b(s)-\b_0 \right\}ds  \right\|\\
&\leq \frac{1}{k_n}\sum_{i=1}^{\lfloor t k_n \rfloor}\left\|  \V_{\b_0}(Z \vert t_i)^{-1/2}\left\{
\E_{\b_0}(Z \vert t_i)
-
\E_{\b(t_i)}(Z \vert t_i) -\V_{\b_0}(Z \vert t_i)\left\{\b(t_i)-\b_0\right\}\right\}\right\| \\
&\quad + \frac{1}{k_n}\sum_{i=1}^{\lfloor t k_n \rfloor} \left\| \left( \V_{\b_0}(Z \vert t_i)^{1/2} - \S^{1/2}\right)\left\{ \b(t_i)-\b_0 \right\} 
\right\|  \\
&\quad +   \left\|\S^{1/2} \left( \dfrac{1}{k_n}\sum_{i=1}^{\lfloor t k_n\rfloor}\b(t_i)-\int_0^t\b(s)ds + \b_0 \left(\dfrac{\lfloor t k_n\rfloor}{k_n}-t \right) \right)\right\| \\
&\leq 
\dfrac{pM_n}{2k_n}\sum_{i=1}^{\lfloor t k_n \rfloor}\left\|  \V_{\b_0}(Z \vert t_i)^{-1/2} \right\| \left\| \b(t_i)-\b_0 \right\|^2 +\frac{p}{k_n}\sum_{i=1}^{\lfloor t k_n \rfloor} \left\|  \V_{\b_0}(Z \vert t_i)^{1/2}-\S^{1/2} \right\| \left\| \b(t_i)-\b_0 \right\| \\
&\quad + p \left\|\S^{1/2} \right\| \left\|\dfrac{1}{k_n}\sum_{i=1}^{\lfloor t k_n\rfloor}\b(t_i)-\int_0^t\b(s)ds\right\|
+\left|\dfrac{\lfloor t k_n\rfloor}{k_n}-t \right| \left\|\S^{1/2} \b_0  \right\| 
\\
&\leq  \frac{\lfloor t k_n \rfloor}{k_n}\dfrac{pM_n}{2}\max_{i=1,\dots,\lfloor t k_n \rfloor}\left\|  \V_{\b_0}(Z \vert t_i)^{-1/2} \right\|
\max_{i=1,\dots,\lfloor t k_n \rfloor}\left\| \b(t_i)-\b_0 \right\|^2  \\
&\quad + p \frac{\lfloor t k_n \rfloor}{k_n} \max_{i=1,\dots,\lfloor t k_n \rfloor}\left\| \b(t_i)-\b_0 \right\| \max_{i=1,\dots,\lfloor t k_n \rfloor} \left\|  \V_{\b_0}(Z \vert t_i)^{1/2}-\S^{1/2} \right\| \\
&\quad + p\left\|\S^{1/2} \right\|\max_{l=1,\dots,p}\left|\dfrac{1}{k_n}\sum_{i=1}^{\lfloor t k_n\rfloor}\b(t_i)_l-\int_0^t\b(s)_lds \right|+p\left\vert \dfrac{\lfloor t k_n \rfloor}{k_n} -t \right\vert \left\|\S^{1/2} \right\|   \left\| \b_0   \right\|. 
\end{align*}

This norm converges to 0 in probability as $n$ increases without bound by the boundedness of the variances (Assumptions \ref{point1_multiv} and \ref{point2_multiv}), their convergence to $\S$ (Assumption \ref{homosced_multiv}) and the convergence to 0 of $M_n$ as $n\rightarrow\infty$. \hfill $\square$

%%%%%%%%%%%%%%%%%%%%%%%%%%%%%%%%%%%%%%%%%%%%%%%%%%%%%%%%%%%%%%%%%%%%%%%%%%%%%%%%%%%%%%%%%%%
%
%
\section{Proof of Proposition \ref{confband}}
\label{appendix_confband}
Let $t\in[0,1]$.
By Theorem \ref{thNPH_multi} and since $\hat{\boldsymbol{ \Sigma}}^{-1/2}$ is a consistent estimator of $ \boldsymbol{ \Sigma}^{-1/2}$, in addition to Slutsky's lemma, we have $\ 
\hat{ \boldsymbol{ \Sigma}}^{-1/2}\left(\U^*(\b_0,t) -t\U^*(\b_0,1)\right) \overset{D}{\underset{n\rightarrow + \infty}{\longrightarrow}}
 \boldsymbol{ \Sigma}^{-1/2}\mathbf{B}_p(t).$ Therefore, 
\begin{equation*}
\left\| \hat{\Sigma}_{\cdot,i}^{-1/2} \right\|_2^{-1}\hat{ \boldsymbol{ \Sigma}}^{-1/2}\left(\U^*(\b_0,t) -t\U^*(\b_0,1)\right) \overset{D}{\underset{n\rightarrow + \infty}{\longrightarrow}}
B(t),
\end{equation*} 
 where $B$ is a Brownian Bridge. The result follow from the knowledge of the limit distribution of the supremum of the absolute value of a Brownian bridge, which is the Kolmogorov distribution. \hfill $\square$

%%%%%%%%%%%%%%%%%%%%%%%%%%%%%%%%%%%%%%%%%%%%%%%%%%%%%%%%%%%%%%%%%%%%%%%%%%%%%%%%%%%%%%%%%%
\section{Proof of Theorem \ref{theoremR2}}
\label{appendix_R2}

We consider first the univariate case, in which $p=1$.
Let us study the numerator of the $R^2$ coefficient defined in equation (\ref{R2def}). We have
\begin{align} \label{eq1}
\dfrac{1}{k_n}\overset{k_n}{\underset{i=1}{\sum}}& \left( \mathcal{Z}(t_i)- \E_{\alpha(t_i)}( Z \mid t_i)\right)^2  \nonumber \\
&=\dfrac{1}{k_n}\overset{k_n}{\underset{i=1}{\sum}} \left( \mathcal{Z}(t_i)- \E_{\beta(t_i)}( Z \mid t_i)\right)^2 +\dfrac{1}{k_n}\overset{k_n}{\underset{i=1}{\sum}} \left( \E_{\beta(t_i)}(Z \mid t_i) - \E_{\alpha(t_i)}( Z \mid t_i)\right)^2\nonumber \\
&+\dfrac{2}{k_n} \overset{k_n}{\underset{i=1}{\sum}} \left( \mathcal{Z}(t_i)- \E_{\beta(t_i)}( Z \mid t_i)\right)\left( \E_{\beta(t_i)}(Z \mid t_i) - \E_{\alpha(t_i)}( Z \mid t_i)\right).
\end{align}
Let us study the right--hand side of equation (\ref{eq1}).
Recall that the random variables $ \mathcal{Z}(t_i)- \E_{\beta(t_i)}( Z \mid t_i)$ are independent for $i=1,\dots,k_n$, that $Z(t)$ admits a moment of order $4$ and  $E\left( \left( \mathcal{Z}(t_i)- \E_{\beta(t_i)}( Z \mid t_i)\right)^2\right)=E\left( \V_{\beta(t_i)}( Z \mid t_i)\right)$. Therefore,
\begin{equation*}
\lim_{n \rightarrow \infty}\sum_{i=1}^{k_n}\dfrac{1}{i^2}E\left( \left\{  \mathcal{Z}(t_i)- \E_{\beta(t_i)}( Z \mid t_i) \right\}^4\right) <\infty.
\end{equation*}
Markov's law of large numbers for independent and non--identically distributed random variables imply that
\begin{equation}
\dfrac{1}{k_n} \overset{k_n}{\underset{i=1}{\sum}} \left( \mathcal{Z}(t_i)- \E_{\beta(t_i)}( Z \mid t_i)\right)^2 - \dfrac{1}{k_n} \overset{k_n}{\underset{i=1}{\sum}} E \left(\V_{\beta(t_i)}( Z \mid t_i) \right) \overset{\mathbb{P}}{\underset{n \rightarrow \infty} \longrightarrow} 0.
\end{equation}
By Lemma 1 of \citet{Chauvel2014}, we have
\begin{equation} \label{CVPvar}
\dfrac{1}{k_n} \overset{k_n}{\underset{i=1}{\sum}} \V_{\beta(t_i)}( Z \mid t_i)\overset{\mathbb{P}}{\underset{n \rightarrow \infty} \longrightarrow} \int_0^1 v(\beta(t),t)dt.
\end{equation}
Conditional empirical variances are almost surely bounded implying that

 \begin{equation*}
 \dfrac{1}{k_n} \overset{k_n}{\underset{i=1}{\sum}} \left( \mathcal{Z}(t_i)- \E_{\beta(t_i)}( Z \mid t_i)\right)^2\overset{\mathbb{P}}{\underset{n \rightarrow \infty} \longrightarrow} \int_0^1 v(\beta(t),t)dt.
 \end{equation*}
The convergence of the second term of equation (\ref{eq1}) is again obtained by Lemma 1 of \citet{Chauvel2014}: 
\begin{equation*}
\dfrac{1}{k_n} \overset{k_n}{\underset{i=1}{\sum}} \left( \E_{\beta(t_i)}(Z \mid t_i) - \E_{\alpha(t_i)}( Z \mid t_i)\right)^2\overset{\mathbb{P}}{\underset{n \rightarrow \infty} \longrightarrow} \int_0^1 \left(e(\alpha(t),t)-e(\beta(t),t)\right)^2dt.
 \end{equation*}
 Finally, the last term of equation (\ref{eq1}) converges in probability to 0 when $n\rightarrow\infty$ by Markov's law of large numbers. Thus,
\begin{align} \label{CVR2univ}
\lim_{n\rightarrow \infty}R^2\left(\alpha(t) \right)=1-\dfrac{\int_0^1 v(\beta(t),t)dt+\int_0^1 \left(e(\alpha(t),t)-e(\beta(t),t)\right)^2dt}{\int_0^1 v(\beta(t),t)dt+\int_0^1 \left(e(0,t)-e(\beta(t),t)\right)^2dt},
\end{align} 
and $\lim_{n\rightarrow \infty}R^2 $ reaches its maximum in $\beta(t)$.\newline

For the multivariate case ($p>1$), similar arguments lead to the limit
\begin{equation} \label{CVR2multiv}
\lim_{n\rightarrow \infty}R^2\left(\a(t) \right)=1-\dfrac{\int_0^1 \a(t)^Tv(\b(t),t)\a(t)dt+\int_0^1 \left( \a(t)^T\left\{e(\b(t),t)-e(\a(t),t)\right\}\right)^2dt}{\int_0^1 \a(t)^Tv(\b(t),t)\a(t)dt+\int_0^1 \left( \a(t)^T\left\{e(\b(t),t)-e({\bf 0},t)\right\}\right)^2dt}.
\end{equation}
Finally, 
$\left| R^2( \b(t))- R^2(\hat \b(t) )\right|\ \overset{a.s.}{\underset{n\rightarrow \infty}{\longrightarrow}} 0,\ $ as $\hat \b(t)$ is a consistent estimator of $\beta(t)$. \hfill $\square$

%%%%%%%%%%%%%%%%%%%%%%%%%%%%%%%%%%%%%%%%%%%%%%%%%%%%%%%%%%%%%%%%%%%%%%%%%%%%%%%%%%%%%%%%%%
\section{Proof of Theorem \ref{theoremR2L2}}
\label{proofR2L2}

%%%%%%%%%%%%%%%%%%%%%%%%%ù
Let $p=1$, $\alpha(t)\in\mathcal{B}$ and assume that \ref{point1_multiv} and \ref{point2_multiv} are verified. A Taylor series expansion of $e\left(\alpha(t),t\right) $ in equation (\ref{CVR2univ}) gives
\begin{align*}
\lim_{n\rightarrow \infty}R^2\left(\alpha(t) \right)=1-\dfrac{\int_0^1 v(\beta(t),t)dt+\int_0^1 \left(\alpha(t)-\beta(t)\right)^2v\left(c(t),t\right)^2dt}{\int_0^1 v(\beta(t),t)dt+\int_0^1 \left(e(0,t)-e(\beta(t),t)\right)^2dt},
\end{align*} 
where $c(t)$ lies between $\alpha(t)$ and $\beta(t)$. 
Therefore, minimizing $ \lim_{n\rightarrow \infty}R^2\left(\alpha(t) \right)$ in $\alpha(t)$ reduces to minimize $ \int_0^1 \left(\alpha(t)-\beta(t)\right)^2v\left(c(t),t\right)^2dt$.
\hfill $\square$

%%%%%%%%%%%%%%%%%%%%%%%%%%%%%%%%%%%%%%%%%%%%%%
%%%%%%%%%%%%%%%%%%%%%%%%%%%%%%%%%%%%%%%%%%%%%%

\section*{Supplementary Material}

Consider the setting of the proof of Theorem 1.
Let us verify that the hypotheses of the functional central limit theorem for martingale differences of \citet{Helland1982} are satisfied. Let $t \in [0,1]$ and $l,m=1,\dots,p,$ with $l\neq m$. Denote $\e_l$ the $l$th vector of the standard basis of $\R^p$: all of its elements are null except for its $l$th element which equals $1$. Then, 
\begin{equation*}
\xi_{i,k_n}^l=\e_l^T\,\bxi_{i,k_n}=\bxi_{i,k_n}^T\,\e_l \in \R.
\end{equation*}

\begin{enumerate}[A.]
\item (Martingale difference array.) 
Using the inclusions of the $\sigma$-algebras $\mathcal{F}_{t_{i-1}}\subset \mathcal{F}_{t_{i}} $ and the centering of the increments, we have
\begin{equation*}
E_{\b(t_{i-1})}\left(\left.\xi_{i,k_n}^l \right|t_{i-1}\right)
=E_{\b(t_{i-1})}\left(\left. E_{\b(t_i)}\left(\xi_{i,k_n}^l|t_i\right)\right|t_{i-1}\right)=0.
\end{equation*}
%%%%%%%%%%%%%%%%%%%%%%%%%%%%%%%%%%%%%%%%%%%%%%
\item (Uncorrelatedness.) 
Notice that 
\begin{align*}
E_{\b(t_{i})}(\xi_{i,k_n}^l\xi_{i,k_n}^m  |t_i)
&=\e_l^T\,\E_{\b(t_{i})}(\bxi_{i,k_n}\bxi_{i,k_n}^T  |t_i)\, \e_m
\\
&=\dfrac{1}{k_n}\e_l^T\,\V_{\b_0}(Z\vert t_i)^{-1/2} 
\V_{\b(t_{i})}(Z \vert t_i )
\V_{\b_0}(Z\vert t_i)^{-1/2} \,\e_m 
\end{align*}
Therefore, using the inclusion of sigma-algebras,
\begin{align*}
&E\left( \left\vert\sum_{i=1}^{ \lfloor t k_n \rfloor }E_{\b(t_{i-1})}\left( \left.\xi_{i,k_n}^l\xi_{i,k_n}^m  \right\vert t_{i-1}\right)
 \right|\right) \\
 &=E\left( \left\vert \dfrac{1}{k_n}\sum_{i=1}^{ \lfloor t k_n \rfloor }E_{\b(t_{i-1})}\left( \left. \e_l^T\, \V_{\b_0}(Z\vert t_i)^{-1/2} 
\V_{\b(t_{i})}(Z \vert t_i )
\V_{\b_0}(Z\vert t_i)^{-1/2} \,\e_m \right\vert t_{i-1}\right)  
 \right|\right) \\
 & \leq \dfrac{1}{k_n}\sum_{i=1}^{ \lfloor t k_n \rfloor } E\left( \left\vert \e_l^T\,\V_{\b_0}(Z\vert t_i)^{-1/2} 
\V_{\b(t_{i})}(Z \vert t_i )
\V_{\b_0}(Z\vert t_i)^{-1/2}  \,\e_m\right|\right)\\
& \leq \dfrac{\lfloor t k_n \rfloor}{k_n} \max_{i=1,\dots,\lfloor k_nt\rfloor} E\left( \left\vert \e_l^T\,\V_{\b_0}(Z\vert t_i)^{-1/2} 
\V_{\b(t_{i})}(Z \vert t_i )
\V_{\b_0}(Z\vert t_i)^{-1/2}  \,\e_m\right|\right).
\end{align*}
By assumption C 
and the continuous mapping theorem for matrices and vectors, 
\begin{equation*}
\e_l^T\,\V_{\b_0}(Z\vert t_i)^{-1/2} 
\V_{\b(t_{i})}(Z \vert t_i )
\V_{\b_0}(Z\vert t_i)^{-1/2}  \,\e_m \overset{P}{\underset{n\rightarrow \infty}{\longrightarrow}} \e_l^T\,\S^{-1/2}\S\S^{-1/2}\,\e_m=0.
\end{equation*}
This convergence is also a convergence in mean by the almost sure boundedness of each quantity. Thus, $E_{\b(t_{i-1})}(\xi_{i,k_n}^l\xi_{i,k_n}^m  |t_{i-1}) \overset{L^1}{\underset{n\rightarrow \infty}{\longrightarrow}} 0.$
%%%%%%%%%%%%%%%%%%%%%%%%%%%%%%%%%%%%%%%%%%%%%%%%%%%%%%
\item (Variance.) Denote $\mathbf{I}_p$ the identity matrix of dimension $p\times p.$ The same arguments leads us to the following equality
\begin{align*}
&\sum_{i=1}^{ \lfloor t k_n \rfloor }E_{\b(t_{i-1})}\left( \left.\left(\xi_{i,k_n}^l\right)^2  \right\vert t_{i-1}\right)-t \\
&=
\dfrac{1}{k_n} \sum_{i=1}^{ \lfloor t k_n \rfloor }E_{\b(t_{i-1})}\left( \left.   \e_l^T\,\left\{\V_{\b_0}(Z\vert t_i)^{-1/2} 
\V_{\b(t_{i})}(Z \vert t_i )
\V_{\b_0}(Z\vert t_i)^{-1/2} -\mathbf{I}_p\right\}  \e_l \right\vert t_{i-1}\right)+ \dfrac{\lfloor k_nt \rfloor }{k_n} -t.
\end{align*} 
Thus,
\begin{align*}
&E \left(\left\vert \sum_{i=1}^{ \lfloor t k_n \rfloor }E_{\b(t_{i-1})}\left( \left.\left(\xi_{i,k_n}^l\right)^2  \right\vert t_{i-1}\right)-t \right\vert \right)\\
 &\leq  \dfrac{1}{k_n} \sum_{i=1}^{ \lfloor t k_n \rfloor }E \left.\bigg( 
\right\vert 
\e_l^T\, \left\{\V_{\b_0}(Z\vert t_i)^{-1/2} 
\V_{\b(t_{i})}(Z \vert t_i )
\V_{\b_0}(Z\vert t_i)^{-1/2} -\mathbf{I}_p\right\} 
 \e_l \left\vert \bigg)\right. +\left\vert \dfrac{\lfloor k_nt \rfloor }{k_n} -t \right\vert .
\end{align*}
Again, assumption C
, the continuous mapping theorem and the almost sure boundedness of the variances imply
\begin{align*}
\e_l^T\, \left\{\V_{\b_0}(Z\vert t_i)^{-1/2} \right. &\left.\V_{\b(t_{i})}(Z \vert t_i )
\V_{\b_0}(Z\vert t_i)^{-1/2} -\mathbf{I}_p\right\}  \e_l %\\
%&
 \overset{L^1}{\underset{n\rightarrow \infty}{\longrightarrow}} \e_l^T\,\left\{\S^{-1/2}\S\S^{-1/2}-I\right\}\e_l=0 .
\end{align*}
Therefore, $ \sum_{i=1}^{ \lfloor t k_n \rfloor }E_{\b(t_{i-1})}\left( \left.\left(\xi_{i,k_n}^l\right)^2  \right\vert t_{i-1}\right) \overset{L^1}{\underset{n\rightarrow \infty}{\longrightarrow}} t .$
%%%%%%%%%%%%%%%%%%%%%%%%%%%%%%%%%%%%%%%%%%%%%%%%%%%%%%%%%%%
\item (Lyapunov condition.) By the boundedness of the increments $\left(\xi_{i,k_n}^l\right)_i$, there exists a constant $C \in [0,+\infty[$ such that for all $i=1,\dots,k_n,$ $E_{\b(t_{i})}\left(\left. \left\vert \xi_{i,k_n}^l\right\vert^3  \right\vert t_{i}  \right) \leq C$ almost surely. Thus,
\begin{align*}
& \sum_{i=1}^{\lfloor k_n t \rfloor} E_{\b(t_{i-1})}\left(\left. \left\vert \xi_{i,k_n}^l \right\vert^3  \right\vert t_{i-1} \right)
\leq \dfrac{1}{k_n^{3/2}}\sum_{i=1}^{\lfloor k_n t \rfloor}E_{\b(t_{i-1})}\left(\left.  
E_{\b(t_{i})}\left(\left. \left\vert \xi_{i,k_n}^l\right\vert^3  \right\vert t_{i}  \right)
\right\vert t_{i-1} \right) 
\leq  \dfrac{\lfloor k_n t \rfloor}{k_n^{3/2}}C.
\end{align*}
Hence, $ \sum_{i=1}^{\lfloor k_n t \rfloor} E_{\b(t_{i-1})}\left(\left. \left\vert \xi_{i,k_n}^l \right\vert ^3  \right\vert t_{i-1} \right) \overset{P}{\underset{n\rightarrow \infty}{\longrightarrow}}0.$
\end{enumerate}
As a conclusion, all hypotheses of Helland's multivariate functional central limit theorem are gathered and $\X_n$ converges weakly to a multivariate Wiener process $\mathbf{W}_p$ as $n$ increases without bound.

\singlespacing
\begin{small}
%\bibliography{../biblio}
\bibliography{biblio}
\bibliographystyle{apalike}
\end{small}

\end{document}